\documentclass[reprint,
showpacs,preprintnumbers,amsmath,amssymb, aip,jcp,nolongbibliography]{revtex4-1}
\usepackage{fancyhdr}
\usepackage[usenames,dvipsnames]{color}
\usepackage{graphicx}
\usepackage{dcolumn}
\usepackage{bm}
 \usepackage{float}
\usepackage{natbib}
\newcommand{\kB}{k_{\mathrm{B}} T}

\definecolor{light-gray}{gray}{0.45}

\begin{document}

\preprint{1}

\title{The Putative Liquid-Liquid Transition is a Liquid-Solid Transition in Atomistic Models of Water, Part II }

\author{David T. Limmer}
\author{David Chandler}
 \email{chandler@berkeley.edu}
\affiliation{%
Department of Chemistry, University of California, Berkeley, California 94720
}%

\date{\today}
\begin{abstract}
This paper extends our earlier studies of free energy functions of density and crystalline order parameters for models of supercooled water, which allows us to examine the possibility of two distinct metastable liquid phases [{\it J. Chem. Phys.} \textbf{135}, 134503 (2011) and arXiv:1107.0337v2].    Low-temperature reversible free energy surfaces of several different atomistic models are computed:  mW water, TIP4P/2005 water, SW silicon and ST2 water, the last of these comparing three different treatments of long-ranged forces.  In each case, we show that there is one stable or metastable liquid phase, and there is an ice-like crystal phase. The time scales for crystallization in these systems far exceed those of structural relaxation in the supercooled metastable liquid.  We show how this wide separation in time scales 
produces an illusion of a low-temperature liquid-liquid transition. The phenomenon suggesting metastability of two distinct liquid phases is actually coarsening of the ordered ice-like phase, which we elucidate using both analytical theory and computer simulation.  For the latter, we describe robust methods for computing reversible free energy surfaces, and we consider effects of electrostatic boundary conditions.  We show that sensible alterations of models and boundary conditions produce no qualitative changes in low-temperature phase behaviors of these systems, only marginal changes in equations of state.  On the other hand, we show that altering sampling time scales can produce large and qualitative non-equilibrium effects.  Recent reports of evidence of a liquid-liquid critical point in computer simulations of supercooled water are considered in this light.

\end{abstract}

\maketitle

\section{\label{sec:level1}Introduction}
This is our second paper examining whether molecular simulation provides support for the  hypothesis that supercooled water possesses two distinct liquid phases with a reversible coexistence line ending at a critical point.~\cite{Poole:1992p324}  In the first (Paper I),\cite{Limmer:2011p134503} we described our results for pertinent free energy functions of three different models: the mW model of water, \cite{Molinero:2009p4008} a variant of the ST2 model for water, \cite{Stillinger:1974p1545} and the Stillinger-Weber model for Si.~\cite{Stillinger:1985p5262}  Each of these models has an equilibrium liquid phase with pair distributions and thermodynamic anomalies like those of water, and each has an equilibrium phase transition like that of water-ice freezing.  For each, others have claimed numerical evidence of liquid-liquid transitions at supercooled conditions,~\cite{Poole:1992p324,sastry2003liquid,brovchenko2003multiple, Kumar:2005p2095, beaucage2005liquid, liu2009low,sciortino2011study,moore2009growing,xu2011there,liu2009low,vasisht2011liquid,sciortino2011study,liu2012liquid, kesselring2012nanoscale, kesselring2012liquid, kesselring2013finite, poole2013free} but our calculations described in Paper I reveal no such behavior.  Rather, for each system we found that there can be at most one stable or metastable liquid phase, and this liquid can coexist with crystalline ice.  Here, we establish that results contrary to our findings derive from lack of equilibration of slow fluctuations in long range order.  We also present new calculations for several additional models reaching consistent conclusions in each case.  Specifically, for computer-simulation models of water that exhibit liquid and ice-like phases, there is no liquid-liquid transition, but there is non-equilibrium coarsening of ice that others have misinterpreted as evidence of a liquid-liquid transition.

\begin{figure}
\begin{center}
\includegraphics[width= 8.5cm]{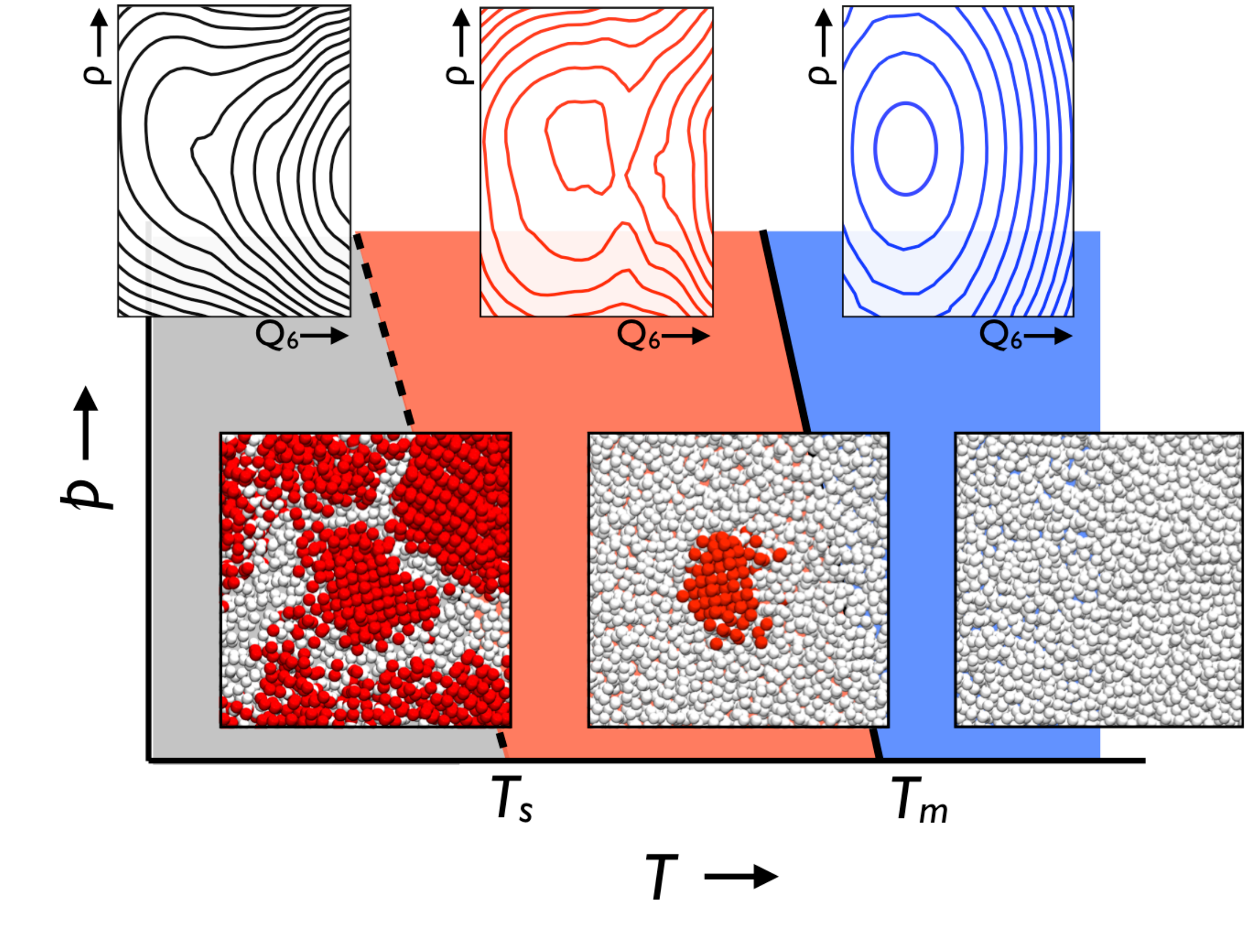}
\caption{Phase diagram, free energy surfaces and typical configurations of cold water.  Typical of all systems considered in this paper, the specific pictures render results from molecular simulations of one particular model.  Quantitative scales of temperature, pressure and free energy depend upon simulation model, and these scales are omitted here because this figure serves a qualitative purpose only.  For experimental water, the phase diagram covers pressures $p$ ranging from 1~bar to 3~kbar, and temperatures $T$ ranging from 150~K to 300~K.  $T_\mathrm{m}$ and $T_\mathrm{s}$ stand for temperatures at the melting and liquid-stability lines, respectively.  Blue region in the $p$-$T$ plane is where liquid is stable, red region is where liquid is metastable with respect to ice I, and grey region is where liquid is unstable (i.e., it is the liquid's ``no-man's land'').  Corresponding free energy surfaces are shown above as functions of density $\rho$ and crystal-order variable $Q_6$.  Corresponding molecular configurations shown below are cuts through a simulation box at the ends of trajectories that are initiated in a liquid configuration and that run for times shorter than required to crystalize the entire sample.  Molecules located in crystal-like regions are colored red.}
\label{Fig1}
\end{center}
\end{figure}

Figure \ref{Fig1} shows the relevant part of water's phase diagram and corresponding free energy surfaces.   The liquid is the stable equilibrium phase for temperatures above the melting temperature, i.e.,  $T>T_\mathrm{m}$, and it is unstable below a stability temperature, i.e., $T< T_\mathrm{s}$.  In between, for $T_\mathrm{s}<T<T_\mathrm{m}$, the liquid is metastable with respect to crystal ice.  Throughout much of that intermediate region, structural reorganization of water is slow, and it becomes slower in a super-Arrhenius fashion as temperature is lowered.~\cite{angell:1983p593}  This sluggishness can present problems for straightforward molecular simulation, as noted below, but it is not so sluggish to prevent certain crystallization of water when the liquid is cooled close to or below $T_\mathrm{s}$.  Coarsening of water in that regime occurs on the time scale of microseconds -- fast for experiment, but slow for simulation.~\cite{koop:2000p611}  All speculations on the existence of a liquid-liquid phase transition in water locate that transition near or below $T_\mathrm{s}$, the so-called ``no-man's land'' for liquid water.  As such, it is difficult for experiments to prove or disprove the liquid-liquid hypothesis.  It is left to simulation, which can reversibly control crystallization, to see if such an idea could be correct within the purview of statistical mechanics for plausible models of water or water-like systems.

Calculations of free energy functions of relevant order parameters are required when using simulation to establish phase behavior.~\cite{frenkel2001understanding}  As noted, such calculations can be difficult, especially for supercooled water because fluctuations in this regime are collective and slow.  To address this difficulty and sort out the phase behavior of supercooled water, we have found it convenient to consider two order parameters.  One is molecular density, $\rho$, that distinguishes different amorphous phases.  The other can be the Steinhardt-Nelson-Ronchetti $Q_6$ that distinguishes an amorphous phase from a symmetry-broken crystalline phase.~\cite{steinhardt1983bond}$^,$ \footnote{Formulas for computing $Q_6$ are Eqs. (1) - (3) in Ref.~\onlinecite{Limmer:2011p134503}}  The two variables fluctuate on significantly different time scales.  For example, the liquid structural relaxation time (i.e., the equilibration time for $\rho$) around $T \approx T_\mathrm{s}$ is of order $10^{-8}\,\text{s}$ or shorter, whereas the relevant equilibration time for $Q_6$ in this regime is the time to form a crystal, $10^{-6}\,\text{s}$ or longer.  This wide separation of time scales is typical of systems undergoing crystallization transitions.~\cite{debenedetti1996metastable}  In the case of water, we will see in this paper that it is a principal source of confusion in simulation studies that claim evidence for a liquid-liquid phase transition.

To foreshadow this point, consider the equilibrium joint distribution function for the order parameters, $P(\rho, Q_6)$.  It is related to the free energy (or reversible work) surface for these variables in the usual way:  $F(\rho, Q_6) = -k_\mathrm{B}T \ln P(\rho, Q_6)$, where $k_\mathrm{B}$ is Boltzmann's constant.  This is the free energy function illustrated in Fig.~\ref{Fig1}.  Over time scales large compared to those of liquid relaxation but possibly not large compared to those of crystal formation, the joint distribution is in general a non-equilibrium distribution,
\begin{equation}
\label{eq:factorization1}
P_\mathrm{ne}(\rho, Q_6, t) = P(\rho | Q_6) \, P_\mathrm{ne}(Q_6, t)\,,
\end{equation}
where $P(\rho | Q_6)$ is the equilibrium distribution for $\rho$ given a specific value for $Q_6$, and $P_\mathrm{ne}(Q_6, t)$ is the non-equilibrium distribution for $Q_6$.  The non-equilibrium distribution depends upon the protocol with which the system is prepared, and its time dependence is irreversible.  For large enough $t$, presuming ergodicity, $P_\mathrm{ne}(Q_6,t)$ approaches the equilibrium $P(Q_6)$.  But this limit can require simulation times thousands of times longer than those needed to equilibrate $\rho$.  Not accounting for this behavior can give the illusion of a reversible polyamorphism because the non-equilibrium free energy, $- k_\mathrm{B} T \ln [P(\rho | Q_6) \, P_\mathrm{ne}(Q_6, t)]$, can have a low-$Q_6$ basin for times shorter than those required for $Q_6$ to diffuse towards its equilibrium crystal value at high $Q_6$.  

This possibility, which we refer to as ``artificial polyamorphism,'' can be appreciated by comparing the free energy surfaces shown in Fig.~\ref{Fig1}.  In particular, imagine studying the system on time scales where $Q_6$ can diffuse over no more than the left halves of the pictured free-energy panels.  $P_\mathrm{ne}(Q_6, t)$ would then be peaked at a low value of $Q_6$, even for cases where a high $Q_6$ value would be the correct equilibrium value.  Thus, if $Q_6$ is limited in this way to small values, the low-temperature (i.e., left-most) panel would then yield a pseudo free energy, $- k_\mathrm{B} T \ln P_\mathrm{ne}(\rho, Q_6, t)$, with an illusory ``amorphous basin'' at a density lower than that of the  metastable liquid.  This irreversible behavior was discussed in the Supplement to Paper I.\cite{Limmer:2011arxiv}  Specifically, we showed that for liquid water at pressures and temperatures in or close to ``no man's land,'' small values of $Q_6$ will survive while the crystal phase begins to coarsen.  The bottom left of Fig.~\ref{Fig1} shows a configuration of water in that regime.  

Section II of the current paper provides a quantitative theoretical analysis of this behavior.  It shows specifically how the polyamorphism of Refs.~\onlinecite{liu2009low,sciortino2011study,liu2012liquid, kesselring2012nanoscale,poole2013free} is an irreversible effect reflecting the time-scale separation between fluctuations in $\rho$ and fluctuations in $Q_6$.  During the time of coarsening, the faster order parameter, $\rho$, fluctuates between typical crystal and liquid values.   References \onlinecite{kesselring2012nanoscale,liu2012liquid} report this type of behavior, which they call ``phase flipping.''   On the time scale of the flipping, the drift in $P_\mathrm{ne}(Q_6,t)$ can be almost imperceptible, but drift it does. The authors of Refs.~\onlinecite{kesselring2012nanoscale,liu2012liquid} describe the flipping as evidence of a second metastable liquid.  The analysis of Section II shows that this flipping is not a consequence of such metastability, but rather the coarsening of the crystal from the unstable or nearly unstable liquid, occurring steadily and irreversibly on a time scale long compared to those considered in Refs.~\onlinecite{kesselring2012nanoscale,liu2012liquid}.   This finding could already be anticipated from the Supplement to Paper I \cite{Limmer:2011arxiv} and from Moore and Molinero's previous study of crystallization of mW water.~\cite{moore2010ice}  

The specific pathways by which simulated models coarsen depend upon system size. 
For example, free energy barriers separating coexisting equilibrium basins scale as $N^{2/3}$, manifesting the presence of an interface.~\cite{chandler1987introduction}  Similarly, with increasing $N$, the thermally accessible widths of the two basins decrease as $N^{-1/2}$, and the width of the barrier grows.~\cite{chandler1987introduction}  Due to the growing barrier height and barrier width, the frequency of spontaneous events carrying a system between stable phases is therefore vanishingly small in the limit of large $N$.
Similarly, in or near the region of liquid instability (i.e., $T$ close to or lower than the crossover temperature $T_\mathrm{s}$), the slope towards the crystal basin will increase in magnitude as $N$ increases.  This behavior will affect the rate of ``phase flipping'' during coarsening, giving the transient impression of a non-equilibrium barrier between low-density and high-density states that changes with $N$.

These finite-size effects are fundamental to the nature of phase transitions.  Establishing the existence of a phase transition requires studying system-size dependence, for example, by computing changes in free energy barriers with respect to changing $N$.  No such computations have yet been performed for putative liquid-liquid transitions in models of water that exhibit water-like structure of the liquid and crystal phases.  To do so requires algorithms that can attend to the collective nature of systems undergoing phase transitions.  Free energy methods are among the tools that are suitable for the task, provided they are combined with trajectory algorithms that are appropriately efficient and reversible.~\cite{pohorille2010good}  

In Section III, we detail how pertinent free energies can be computed for supercooled water, and we consider different variants of the ST2 model as applications.  Juxtaposition of free energy surfaces for three different variants indicates that reasonable changes in electrostatic boundary conditions do not change general phase behaviors. Section IV presents free energy surfaces obtained for other systems: the mW of water, the TIP4P/2005 model of water,\cite{abascal2005general} and the SW model of Si.  In every case, the models are found to exhibit one stable or metastable liquid phase plus ice-like crystal phases.  Coexistence between two distinct liquid phases does not occur.  A summary of our findings is given in Section IV, and Appendices A, B and C present further details and results. 

Reversibility is particularly important to the issues addressed here and in Paper I.  Distinct reversible phases can be interconverted, with properties that are independent of the paths by which they are prepared.  Reversible liquid phases are thus not the same as amorphous solids or glasses.  The former are reversible and ergodic, so their measured stationary behaviors are independent of history. The latter, like high-density or low-density amorphous ices (HDA and LDA), are not ergodic, so their behaviors depend much on history (i.e., preparation protocols).  Observed transitions between HDA and LDA phases,\cite{mishima1985apparently} therefore, are necessarily different than reversible liquid-liquid transitions.  Melting amorphous ice to produce a non-equilibrium liquid that then crystalizes is different too.~\cite{kohl2005liquid}   

Crystallization following the melting of glass \cite{elsaesser2010reversibility} and crystallization following the rapid quench of water into the liquid's ``no-man's land'' \cite{angell2008insights} are much like non-equilibrium dynamics evolving from low to high $Q_6$ on the middle and left free energy surfaces pictured in Fig.~\ref{Fig1}, an observation worthy of future study.  But these interesting non-equilibrium processes and the transitions between different amorphous solids of water are not our focus in this work.  Rather, we are concerned with whether water-like systems when constrained to not freeze can exhibit two distinct liquid phases.  If such reversible polymorphism were possible, these systems could also exhibit a second critical point as Stanley and many of his co-workers have proposed.~\cite{Poole:1992p324,Mishima:1998p2948,Mishima:1998p2948,Stokely:2010p6455,Kumar:2005p2095,Mishima:2000p4162,Poole:1994p673, liu2009low,sciortino2011study,liu2012liquid,poole2013free,kesselring2012nanoscale}  If, instead, reversible molecular simulation models exhibit only ice and one liquid, then the symmetry differences between ice and liquid exclude the possibility of an associated critical point.  
We believe the systematic evidence provided herein and in Paper I indicates that there is only one liquid and no low-temperature critical point. 

\begin{figure*}
\begin{center}
\includegraphics[width= 15cm]{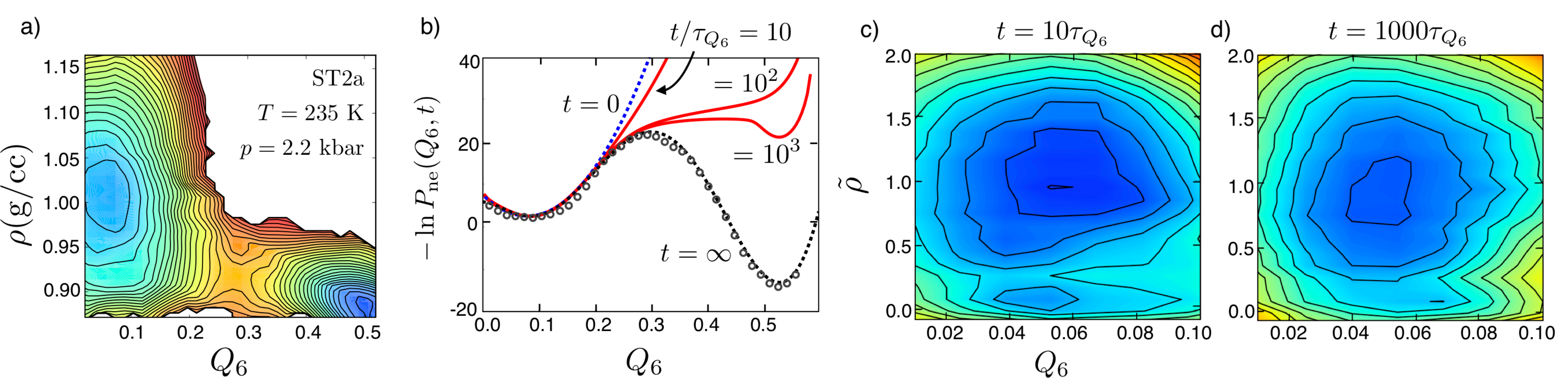}
\caption{Slow relaxation behavior and its consequences for free energy calculations. a)  The reversible free energy surface for 216 molecules with the ST2a potential energy function at temperature $T= 235$~K and  pressure $p=2.2$~kbar.  (See text for definition of the ST2a potential.)  Contour lines are separated by 1.5~$\kB$, and statistical uncertainties are about 1~$\kB$.  b) Negative logarithm of the non-equilibrium distribution for crystal order, $Q_6$, as it relaxes from the liquid state.  It is computed from the Fokker-Planck equation with the free energy surface given in Panel (a) under the assumption that the density, $\rho$, remains at equilibrium with the instantaneous value of $Q_6$.  (c) and (d)  Non-equilibrium pseudo free energy surfaces computed from Eq.~\ref{eq:factorization1} at two intermediate stages of relaxation, $t=10 \,\tau_{Q_6}$ and $t=1,000\, \tau_{Q_6}$.  The unit of time, $\tau_{Q_6}$, is the autocorrelation time for $Q_6$ fluctuations in the liquid basin (i.e., at small $Q_6$).  The reduced density is $\tilde{\rho} = (\rho - \rho_\mathrm{xtl})/ \Delta \rho$, where $\rho_\mathrm{xtl}$ is the mean density of the crystal basin (i.e., at large $Q_6$), and $\Delta \rho$ is the difference between the mean densities of the liquid and crystal basins. Contour lines are separated by 1 $\kB$ and statistical uncertainties are about 1 $\kB$}
\label{Fig2}
\end{center}
\end{figure*}

\begin{table}
  \caption{Separation of timescales for fluctuations in density and long-ranged order.}
  \label{timescales}
   \begin{tabular}{l >{\centering}m{4cm} c }
    \hline
    Model 	& 	$\tau_\rho$ \footnotemark[1]	& 	$\tau_{Q_6}$ \footnotemark[2] \\
    \hline\hline
    mW  	& 10$^3$ MDS\footnotemark[3]	& 10$^5$ MDS\footnotemark[3]	\\ \\
    ST2		& 10$^2$ MCS\footnotemark[4] & 10$^4$ MCS\footnotemark[4]	\\ \\
    ST2		& 10$^6$ MCS\footnotemark[5] & 10$^8$ MCS\footnotemark[5]	\\ \\
    Experiment	& 10$^3$ ps\footnotemark[6] & $>10^6$ ps\footnotemark[7]	\\ \\
    \hline \hline 	       
  \end{tabular}
\footnotetext[1]{Liquid structural relaxation time at temperatures close to $T_\mathrm{s}$.}
\footnotetext[2]{Auto-correlation time for $Q_6$ fluctuations in the liquid at temperatures close to $T_\mathrm{s}$.}
\footnotetext[3]{NPT molecular dynamics steps at $T=200$ K, $p=1$ bar.\cite{Limmer:2011p134503}}
\footnotetext[4]{NPT hybrid Monte Carlo steps at $T=235$ K, $p=2.2$ kbar.  This work; see text for details.}
\footnotetext[5]{NPT single particle Monte Carlo steps at $T=229$ K, $p=2.2$ kbar.\cite{liu2012liquid}}
\footnotetext[6]{Estimated for $T=220$K and $p=1$ bar from analysis in Ref.~\onlinecite{limmer2012phase}.}
\footnotetext[7]{Estimated from the critical cooling rate of $10^6$K/s needed to form amorphous ice.\cite{debenedetti1996metastable}}

\end{table}

\section{Theory of coarsening and artificial polyamorphism in computer simulation of water}
This section provides a quantitative theoretical analysis showing the difficulty in obtaining correct reversible free energy surfaces of supercooled water.  We do so by examining the effects of time-scale separation for dynamics on a reversible free energy surface.  The particular surface we employ is the free energy $F(\rho, Q_6)$ derived in Paper I for a variant of the ST2 model.  This free energy is shown in Fig.~\ref{Fig2}(a).  The methods used to obtain that surface are the subject of the next section, but here we only need to assume that there is such a surface, and that it is qualitatively like the surface shown in Fig.~\ref{Fig2}(a).

Free energy surfaces for several models are derived in Sections III and IV.  The generic features of the free energy surfaces are the same for all the models.  There is a liquid basin at small $Q_6$ and large $\rho$, and a crystal basin at large $Q_6$ and small $\rho$.  For a given temperature, $T$, the relatively stabilities of the basins are controlled by the pressure.  The free energy at pressure $p$ is related to the free energy at pressure $p+\Delta p$ in the usual way, 
\begin{equation}
\label{eq:reweight}
F(\rho, Q_6; p + \Delta p, T) = F(\rho, Q_6; p, T)\,+ \,\Delta p \,N/\rho\,.
\end{equation} 
Accordingly, lowering pressure tips the surface towards the crystal basin, and raising pressure tips it towards the liquid basin.

In cases where the crystal is stable but the system is prepared in the liquid, an irreversible drift towards the crystal will occur.  To the extent that $\rho$ and $Q_6$ are the principal slow variables, this coarsening can be described in terms of motion on the $F(\rho, Q_6)$ surface.  By using this perspective, and specifically by adopting the free energy surface pictured in Fig.~\ref{Fig2}(a) , we illustrate here the generic behavior of early-stage coarsening of ice.   The behavior is not specific to the particular free energy surface.  Rather, it is general consequence of a separation of time sales, where the density $\rho$ equilibrates on time scales that are at least two orders of magnitude shorter than the time scales on which $Q_6$ fluctuates.  Such separations of time scales are typical in natural and computer simulated supercooled water.  See Table~\ref{timescales}.  
	 
\subsection{Time dependence of $P_\mathrm{ne}(Q_6, t)$ }
Due to the separation in time scales, relaxation of $P_\mathrm{ne}(Q_6, t)$ can be accurately estimated by assuming density $\rho$ is always in equilibrium with the current value of $Q_6$.  An appropriate Fokker-Planck equation\cite{chaikin2000principles} is therefore
\begin{equation}\label{eq:FP}
\frac{\partial P_\mathrm{ne}(Q_6,t)}{\partial t} = \\ D \frac{\partial }{\partial Q_6} \left ( \frac{ \partial \beta F(Q_6)}{\partial Q_6} +  \frac{\partial }{\partial Q_6} \right )P_\mathrm{ne}(Q_6,t) \, ,
\end{equation}
where 
\begin{equation}
\beta F(Q_6) = - \ln \left ( \int \mathrm{d}\rho \,e^{-\beta F(\rho,Q_6)} \right ) \,.
\end{equation}
The quantity $F(Q_6)$ is the equilibrium free energy for the crystal-order parameter, $\beta = 1/\kB$, and $D=\langle (\delta Q_6)^2 \rangle/\tau_{Q_6}$ is the diffusion constant projected along the $Q_6$ direction.  The quantity $\langle (\delta Q_6)^2 \rangle \approx 0.01$ is the mean-square fluctuation of $Q_6$ in the liquid basin for the 216-molecule system considered in Fig.~\ref{Fig2}(a). The long-time limit is set by the diffusion constant,
\begin{equation}
\lim_{D t  \rightarrow \infty} P_\mathrm{ne}(Q_6,t) \propto  \exp[-\beta F(Q_6) ]. 
\end{equation}

For quantitative treatments of the ultimate equilibration (i.e., of the final stages of crystal coarsening), Eq.~\ref{eq:FP} could be generalized to include $Q_6$-dependence and memory effects in $D$.  Such generalization could account for the complexity of pathways by which multiple ordered domains reorganize and connect and would be expected to increase the timescales for equilibration. These refinements are unnecessary for the current analysis of early-stage coarsening, where $Q_6$ does not progress far from its values in the liquid.  

\begin{figure*}
\begin{center}
\includegraphics[width= 14cm]{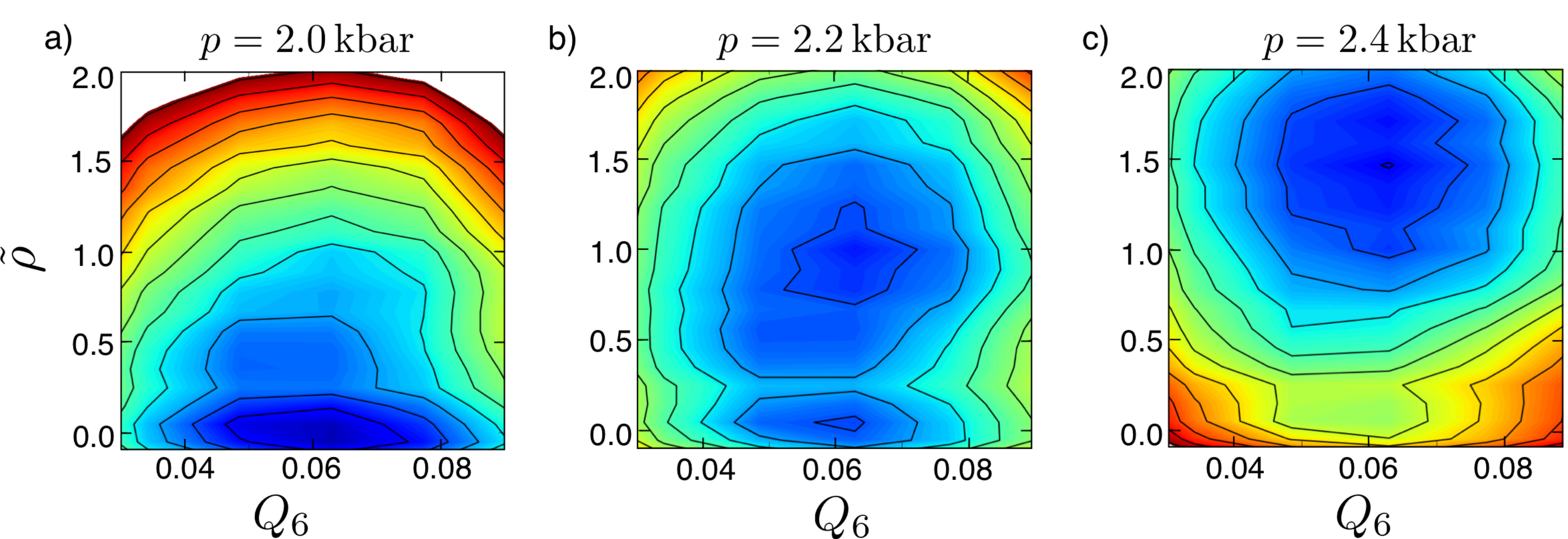}
\caption{Non-equilibrium pseudo free energy surfaces, $F_\mathrm{ne}(\rho, Q_6, t)$, at three different pressures, illustrating how artificial polyamorphism arises as a finite-time effect.  All three surfaces are evaluated by propagating from an initial liquid distribution for a time $t = 10 \tau_{Q_6}$.  Computed as in Fig.~\ref{Fig2}, with the same notation as used in that figure. Contour lines are separated by 1 $\kB$ and statistical uncertainties are about 1 $\kB$. }
\label{Fig3}
\end{center}
\end{figure*}

We have integrated Eq.~\ref{eq:FP} using a first-order finite difference approach with a small enough discretization of $Q_6$ and time 
to ensure numerical stability.~\cite{isaacson1994analysis} Figure \ref{Fig2} (b) shows how the distribution evolves in time from an initial Gaussian distribution centered in the liquid region of $Q_6$. What is notable is that the relaxation to equilibrium takes orders magnitude longer than the basic timescale, $\tau_{Q_6}$. 

With this time evolved probability distribution, we have used Eq.~\ref{eq:factorization1} to estimate a non-equilibrium joint free energy,
\begin{equation}
\label{eq:noneqF}
F_\mathrm{ne}(\rho,Q_6,t) = - \kB \,\ln \left[P(\rho | Q_6)\,P_\mathrm{ne}(Q_6,t)  \right].
\end{equation} 
This pseudo free energy function for two intermediate times is shown in Figs. \ref{Fig2} (c) and (d).  At the first of these intermediate times, $t = 10\, \tau_{Q_6}$, $F_\mathrm{ne}(\rho,Q_6,t)$ exhibits two minima at low values of $Q_6$, and these minima are separated by a small barrier of a few $\kB$. The low density basin is centered at the mean density of the crystal, and the high density basin is centered at the mean density of the liquid.  We use the reduced density variable, $\tilde{\rho} = (\rho - \rho_\mathrm{xtl})/ \Delta \rho$, to emphasize these connections to the crystal and liquid basins.  

This behavior shown in Fig.~\ref{Fig2} (c) is precisely the behavior found in Refs.~\onlinecite{liu2009low,sciortino2011study,liu2012liquid,poole2013free} -- both the bi-stability and the length of time allowed for equilibration.  Those workers find $\tau_{Q_6} \approx 10^8$ Monte Carlo sweeps, and they use $10^{9}$ sweeps to estimate free energies.  The low-density liquid minimum eventually disappears, but the time scale for that to occur is orders of magnitude longer than considered in Refs.~\onlinecite{liu2009low,sciortino2011study,liu2012liquid,poole2013free}.

To further illustrate the connection between the non-equilibrium calculation shown here with the finite-time sampling results of Ref.~\onlinecite{liu2009low,sciortino2011study,liu2012liquid,poole2013free}, Fig.~\ref{Fig3} shows the effects of pressure variation on the pseudo free energy surfaces. 
Upon re-weighting to lower pressure, Fig.~\ref{Fig3} (a), or to higher pressure,  Fig.~\ref{Fig3} (c), one of the disordered minima disappears. The specific dependence on re-weighting and the relative locations of the minima are also consistent with the results of Refs.~\onlinecite{liu2009low,sciortino2011study,liu2012liquid,poole2013free}.  We have thus reproduced the principal results of those papers by identifying the limited time over which the system was allowed to equilibrate.  See for example, Fig. 2 of Ref. \onlinecite{liu2012liquid}, as we have purposely used a similar color code in our Fig.~\ref{Fig3} to emphasize the similarities of our finite-time results with the free energies reported in that paper.

References \onlinecite{liu2009low,sciortino2011study,liu2012liquid,poole2013free} employ slightly different variants of the ST2 model and slightly different temperatures.  In Appendix B, we consider another variant and another temperature to illustrate that the results of our analysis are generic.  The analysis uses the Fokker-Planck equation in order to elucidate the generality of the phenomena.  An explicit simulation calculation establishing the same conclusion is also provided in Appendix B.  

\begin{figure}[b]
\begin{center}
\includegraphics[width= 8.5cm]{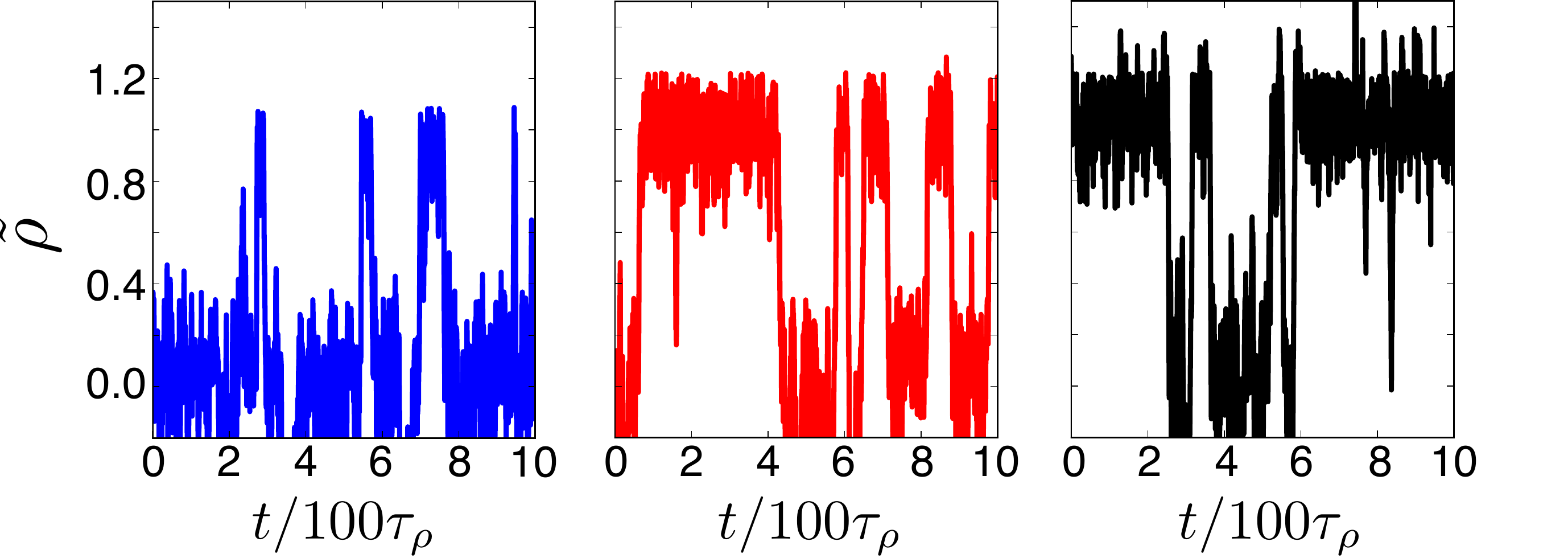}
\caption{Trajectories propagated with over-damped Langevin dynamics on the free energy surface pictured in Fig.~\ref{Fig2}(a).  The trajectories are initiated in the liquid basin and run for insufficient times to pass to $Q_6$ values larger than 0.3.  The trajectories thus illustrate early stages of coarsening in the ST2a model at $T=235$~K and $p=2.2$~kbar.}
\label{Fig4}
\end{center}
\end{figure}

\subsection{Phase flipping}
The time-dependent pseudo free energies illustrated above also shed light on previous reports of phase flipping between two seemingly distinct liquids.\cite{kesselring2012nanoscale,liu2012liquid} In those reports, large transient density fluctuations occur intermittently between smaller amplitude motion while the system is globally liquid-like. This behavior is expected for trajectories of $\rho$ when driven by the pseudo free energy surface graphed in Fig.~\ref{Fig3} (b).  Indeed, Fig.~\ref{Fig4} shows representative trajectories obtained by running over-damped  Langevin dynamics\cite{chaikin2000principles}  on the free energy surface of Fig.~\ref{Fig2}(a) with the time-scale separation $\tau_{Q_6} = 100\, \tau_\rho$.  
The trajectories were initiated in the liquid basin at temperature $T=235$~K. \footnote{The stochastic steps in the Langevin dynamics for $Q_6$ use the diffusion constant $D$ and time scale $\tau_{Q_6}$ specified for the corresponding Fokker-Planck Eq.~\ref{eq:FP}. The stochastic steps for $\rho$ use the diffusion constant  $D_\rho = \left <   (\delta \rho)  ^2 \right > / \tau_\rho$, where $\tau_\rho = 10^{-2} \times \tau_{Q_6}$, and $\left < (\delta \rho)^2 \right > \approx 0.05 (\mathrm{g}/\mathrm{cc})^2$ is the mean-square fluctuation of density in the liquid basin of Fig.~\ref{Fig2}(a). } The trajectories look exactly like those presented as phase flipping in Refs.~\onlinecite{kesselring2012nanoscale,liu2012liquid}.  (The structural relaxation time in those studies is $\tau_\rho \gtrsim 1$~ns.)  

Thus, so-called ``phase flipping'' is not a flipping between distinct liquid phases.  It emerges in a single liquid phase because there exists a separation of timescales between density fluctuations and long range order fluctuations. Large density fluctuations occur as the system is attempting to crystallize, accompanied by the formation of sub-critical nuclei formation that shrink and cause further large density fluctuations. 

This behavior is transient, as the pseudo free energy is ever-changing with time.  An analysis of time series would elucidate this nature of the phenomenon.  Specifically, the mean transition time will show a dependence on trajectory length, reflecting the non-stationarity of the system.  Consequently, the distribution of transition times will deviate from simple Poisson statistics. Those offering phase flipping as evidence for two distinct liquids have not provided a quantitative analysis supporting such statistics. Rather, studies illustrated in Ref. \onlinecite{kesselring2012liquid} catalogue many independent trajectories some of which exhibit anomalously long quiescent periods between transitions lasting several times $100\,\mathrm{ns} \approx 100 \, \tau_\rho$. Data supplied for the ST2 model in Ref.~\onlinecite{kesselring2012liquid} also illustrate that phase flipping occurs at pressures far below a proposed critical pressure, in apparent contradiction to the authors' claims that a critical point exists. The model behavior we discuss is a consequence of liquid-crystal coexistence, which by symmetry must not have a critical point and would explain the insensitivity of this behavior to pressure. 

The only quantitive analysis provided in studies of phase flipping is the calculation of bi-modal density distributions, which are then fit to a universal Ising form, from which supposed critical parameters are extracted.\cite{kesselring2012nanoscale,kesselring2013finite}  The barostat used in those simulations is well known to not reproduce correct equilibrium density fluctuations,\cite{huang2011novel} making such a calculation difficult to interpret. 

\section{Calculation of reversible free energy surfaces }
The previous section emphasizes the significance of a time-scale separation and the pitfalls for simulation that result from it.  In this section, we detail procedures that can overcome the problems inherent in that time-scale separation, and we apply the procedures to the ST2 model to evaluate the liquid-liquid hypothesis in that case. In light of recent speculations,\cite{liu2009low,sciortino2011study,liu2012liquid,poole2013free}  we also analyze how these equilibrium surfaces are affected by changes in model parameters and boundary conditions.

\subsection{Methodology}
We have carried out Monte Carlo simulations with constant number of molecules, $N$, pressure, $p$, and temperature, $T$. This ensemble is appropriate for determining conditions of phase coexistences and is well suited for sampling dense liquid and crystalline states. Density, $\rho = N/V$, fluctuates with $p$ and $N$ fixed because volume, $V$, fluctuates.  The alternative Grand Canonical calculation for sampling global density fluctuations\cite{frenkel2001understanding} can suffer from poor acceptance ratios (e.g., acceptance ratios are reported\cite{liu2009low} to be as low as $10^{-5}$) and can also be ill-defined in applications to crystals.~\cite{frenkel2012simulations} \\ 
\\

\textbf{Trajectory algorithm designed for the task.}
The move set we employ for our Monte Carlo calculations is chosen to mitigate long correlation times expected for supercooled liquids and coarsening crystals. Specifically, to allow for collective reorganizations, we use a hybrid Monte Carlo algorithm\cite{duane1987hybrid}  that propagates an initial configuration with Boltzmann distributed velocities under symplectic, norm preserving, molecular dynamics. For the models with internal degrees of freedom we use the SETTLE integrator while for single-site models we use a velocity Verlet integrator.~\cite{miyamoto1992settle}  The configuration is integrated for a time $n\,\delta t$, where  $\delta t$ is the integration timestep. Each move is accepted with a Metropolis criterion,\cite{frenkel2001understanding} so energy need not be conserved and consequently $\delta t$ need not be small.  

In practice, we generally range from $\delta  t \approx 5-30$ fs and the number of steps in a trial trajectory, $n$, to vary between 1-20 depending of the steepness of the free energy landscape.  The choices are made systematically to minimize correlation times.  Volume moves are used at a ratio of 2 hybrid Monte Carlo moves to 1 trial volume displacement. The relatively large value of $\delta t$ serves to swiftly propagate dynamics over long time scales.  We have found that at supercooled conditions for $N\approx 200$, it significantly reduces correlation times for structural relaxation relative to energy conserving dynamics. For instance, in the case of the ST2 model discussed in detail below, the characteristic structural relaxation times under these moves are between $10^2$ to $10^3$ Monte Carlo steps, depending on the specific value of density and temperature.  In contrast, single particle Monte Carlo moves reported previously\cite{liu2009low,sciortino2011study,liu2012liquid,poole2013free} yield structural relaxation times that are between $10^5$ to $10^7$ Monte Carlo steps, and molecular dynamics\cite{kesselring2012nanoscale} yields structural relaxation times that are between $10^6$ to $10^8$ integration steps. Accounting for the factor $n=\mathcal{O}(10)$, we see that our choice of hybrid Monte Carlo moves is computationally more efficient by 1-3 orders of magnitude over single particle moves and by 2 orders of magnitude over molecular dynamics.  This remarkable speed up must in part reflect the highly non-linear and correlated nature of dynamics at supercooled conditions.  \\

\textbf{Controlled biasing methods.} Fluctuations that result in phase transformations are exponentially rare at conditions of coexistence or modest supercooling. Standard methods of umbrella sampling get around the rare-event problem by adding biasing potentials, $W(x^N)$, to the Hamiltonian in order to enhance occurrences of otherwise improbable fluctuations. Re-weighting configurations correct for the biasing.  Here, $x^N$ stands for a point in the configuration space for the system.  The form of an added energy function can be chosen for convenience, but in order to guarantee the system will reach a stationary state it must be time-independent.  

Free energy methods that do not strictly adhere to this condition, such as meta-dynamics\cite{laio2008metadynamics} and Wang-Landau sampling,\cite{landau2004new} converge only conditionally in the limit that the biasing degree of freedom is the slowest mode or when the change in the biasing term asymptotes to zero.  This issue is particularly relevant for supercooled water because pathways beyond the early stages of coarsening generally involve several slow variables in addition to the global crystal order parameter, $Q_6$.  An incorrect free energy estimate will be obtained if one or more of those slow variables are not controlled in meta-dynamics or Wang-Landau algorithms.  We do not exclude the possibility of inventing adaptive methods that could account for the physical behaviors of supercooled water and be more efficient than the approach we employ, but such adaptive methods are not currently standard.

The order parameters, $\rho$ and $Q_6$, are chosen to distinguish phases of broken symmetries expected to result in water-like models at low temperatures. Our previous study demonstrated that $Q_6$ is a sufficiently sensitive order parameter to distinguish globally ordered from disordered states accompanying a freezing transition.~\cite{Limmer:2011p134503}  This virtue noted, it must also be appreciated that $Q_6$ deviates from its disordered value only after a substantial amount of orientational order has developed in the system.  This fact is demonstrated below and also in the Supplement to Paper I.\cite{Limmer:2011arxiv}  

The umbrella biasing potentials we employ are of the form
\begin{equation}
\label{eq:bias}
W(x^N)= k \left[\rho(x^N) - \rho^* \right]^2 + \kappa \left[Q_6(x^N) - Q_6^* \right ]^2 ,
\end{equation}
where $\rho(x^N)$ and $Q_6(x^N)$ are the density and crystal-symmetry order parameters, respectively for configuration $x^N$.  The biasing potentials, with its force constants $\kappa$ and $k$, keep these order parameters close to their target values $\rho^*$ and $Q_6^*$.  Each pair of target values defines a specific sub-ensemble or so-called ``window'' in configuration space.  After collecting statistics in one window, the window is moved by changing the pair of target values, $\rho^*$ and $Q_6^*$, whereupon statistics in the new window are collected.  The procedure is carried out throughout the $\rho$-$Q_6$ plane, making sure that passage from one region to another is fully reversible, and that adjacent regions have sufficient overlap of statistics to enable further analysis.  

For this purpose, we find that $\kappa$ in the range of 500 to 10,000 $\kB$ and $k$ in the range of 1,000 to 2,000 $\kB$ cm$^6$g$^{-2}$ is satisfactory. Statistics gathered in these biased ensembles are then unweighted and the free energy differences between each ensemble are estimated using MBAR.~\cite{shirts2008statistically}
\begin{figure*}[t]
\begin{center}
\includegraphics[width= 15cm]{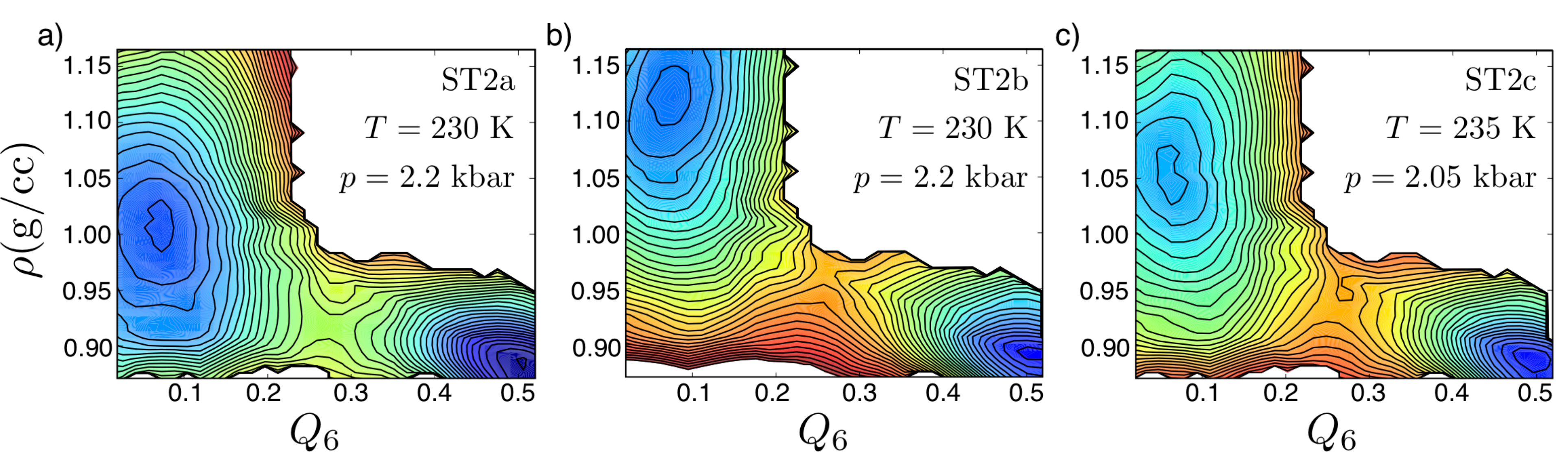}
\caption{Free energy surfaces, $F(\rho,Q_6; p,T )$, for three variants of the ST2 model at temperatures where others report evidence of liquid-liquid coexistence for the ST2 model.  Possibilities of two-phase coexistence require changes in convexity, as re-weighting through Eq.~\ref{eq:reweight} in that case can produce two basins of equal statistical weight.  A coexistence pressure is then the value $p+\Delta p$ at which there is equal statistical weight.  For the three variants considered, the only changes in convexity are associated with coexistence between a liquid (low $Q_6)$ and a crystal (high $Q_6$).  a) Free energy for the ST2a variant at $T=230$ K and $p=2.2$ kbar, with $N=216$. b) Free energy for the ST2b variant at $T=230$ K and $p=2.2$ kbar with $N=216$.  c) Free energy for  the ST2c variant at $T=235$ K and $p=2.05$ kbar with $N=216$.  See text for definitions of the different variants.  Contour lines are separated by $1.5  \kB$ and statistical errors over the surfaces average to less then 1 $\kB$.  Quantitative features will change with system size.  For example, as $N$ grows, the mean value of $Q_6$ in the liquid basin will vanish as $1/N^{1/2}$, while in the crystal basin it will remain finite. }
\label{Fig5}
\end{center}
\end{figure*}

Stitching together data from different sub-ensembles yields the joint probability and thus free energy $F(\rho, Q_6; p, T)$, where $p$ and $T$, respectively, denote the pressure and temperature at which the Monte Carlo trajectory is carried out.  Assuming this function is determined accurately over the relevant range of densities, free energies at other relevant pressures are determined from Eq.~\ref{eq:reweight}, i.e., by straightforward re-weighting.\cite{frenkel2001understanding}$^,$\footnote{It is a simple and well-known point, but unfortunately overlooked by the authors of Refs.\onlinecite{liu2012liquid} and \onlinecite{poole2013free} when they suggest that our earlier Paper I\cite{Limmer:2011p134503} and its Supplement\cite{Limmer:2011arxiv} do not already provide relevant data at the pressures of interest.}

\textbf{Care applied to ensure reversibility.}  In each window, we initially equilibrated for up to 100 structural relaxation times as evaluated for the larger of either the initial liquid density or the equilibrium of bias window density. Production runs were between 50 and 1000 structural relaxation times, depending upon the length of time required to obtain reliable statistics as judged from cumulative averages for $Q_6$ and potential energy.  Here, structural relaxation time refers to the simulation time, $t$, required for mean square fluctuations in structure factors to decay from their initial to 90\% of their relaxed value.

Three different techniques for generating initial conditions were used. Initial seeds were created by cooling an equilibrium liquid initially prepared at $T=330$ K at a rate of 10 K/ns until it reached the target temperature. Seeds from this procedure were biased into different windows in steps between adjacent windows, by gradually changing parameters of the biasing potential $W(x^N)$, Eq.~\ref{eq:bias}, and with re-equilibration runs in between each step.  At high $Q_6$, the crystal that was spontaneously formed using this procedure in all cases was a defected Ice Ic. For second generation seeds, we assumed that the spontaneously formed crystal was the relevant solid phase, so we prepared a perfect Ice Ic configuration, which was used to sample intermediate and high $Q_6$ states as well as bias them into low $Q_6$ regions to sample liquid states. Third generation seeds were obtained by melting an Ice Ic configuration and then using states along the melting trajectory to seed intermediate $Q_6$ windows. These configurations were subsequently biased into the high and low $Q_6$ regions, again by gradually changing the parameters of $W(x^N)$, and again with new re-equilibration runs.  In total, of the order of $10^3$ independent biasing windows are used in the calculation of an individual free energy surface. Specifically, we generated about 800 for each SW and mW free energy surface and about 2500 for each ST2 and TIP4P/2005 model free energy surface.

One issue to keep in mind if one chooses to use parallel tempering and replica exchange~\cite{frenkel2001understanding} is that the shape of the free energy surface changes with temperature.  A liquid basin exists for $T > T_\mathrm{s}$, but it does not exist for $T \lesssim T_\mathrm{s}$.  Moving between replicas that traverse this crossover will produce a transient shadow of the higher temperature liquid basin in the lower temperature replica.  Appendix C illustrates how overlooking this behavior can lead to poor free energy estimates and false impressions of a second liquid basin.  By using a control variable other than temperature, related methods might be designed to increase the relevant diversity of sampled configurations, but the underlying physics that makes a particular variable either applicable or inapplicable must always be considered.   

In principle, free energies can be correctly computed by any number of different methods, provided the procedures are truly reversible.  This property, reversibility, was explicitly checked in our calculations by constructing plots of all two-dimensional histograms and checking for hysteresis. Estimates of errors in free energy differences were made by computing overlaps and gradients of the distributions obtained by various routes. These steps, including bidirectional biasing to and from the crystalline phase and creating many independent realizations of initial conditions, follow standard practices for computing free energies articulated in reviews such as Ref.~\onlinecite{pohorille2010good}. 

\subsection{Results for different variants of the ST2 model}
Figure \ref{Fig5} shows free energy surfaces we have computed for three different versions of the ST2 model, variants that differ only in the manner by which long-ranged forces are computed.  The phase behaviors in each case are similar, with one liquid basin and one crystal basin.  Indeed, the \emph{existence} of singularities in a partition sum is usually not sensitive to subtle changes in potential energy function.  This is true because the existence of a phase transition is mostly dictated by dimensionality and the general form and symmetry of the potential energy function.~\cite{goldenfeld1992lectures}  In contrast, the \emph{locations} of the singularities (e.g., temperatures and pressures of coexistence) are often sensitive to subtle changes.~\cite{smit1991vapor}  Recent reports on variants of the ST2 model~\cite{liu2012liquid,poole2013free} have hypothesized that differences in electrostatic boundary condition and non-electrostatic cutoff parameters can account for why our previously published results~\cite{Limmer:2011p134503} found no liquid-liquid phase transition where others suggest it does exist. The results shown in Fig.~\ref{Fig5} dismiss this hypothesis.

\textbf{The ST2a model.}  Panel (a) is for the model we used in Ref. \onlinecite{Limmer:2011p134503}, but at a temperature slightly lower than that considered in our earlier work.  Our prior reported calculations were for $T=235$~K, whereas Fig. \ref{Fig5}(a) is for $T=230$ K.  The results for $T=230$ K differ very little from those at $T=235$ K.  The model employs a modification of the original ST2 potential for water.\cite{Stillinger:1985p5262}  The modification includes forces from the long ranged electrostatics that were neglected in the original model.  The inclusion uses an Ewald summation with conducting boundary conditions.  The non-electrostatic Lennard-Jones potential is truncated and shifted at 7.5$\mathrm{\AA}$.  This model was referred to as the mST2 model in Ref.~\onlinecite{Limmer:2011p134503}.  Here, we identify it as the ST2a model, and distinguish it from a modified ST2 model with insulating boundary conditions.

\textbf{The ST2b model.}  Panel (b) shows the free energy we have computed at the same temperature as considered in Panel (a), but now with insulating boundary conditions.  In addition, a tail correction has been added to account for the portion of the Lennard-Jones potential neglected in truncation. Panel (b), therefore, shows our results for the equilibrated free energy surface of precisely the variant of the ST2 model used in Refs. \onlinecite{liu2009low,liu2012liquid}.   Here, we identify it as the ST2b model.  As in Panel (a), the surface exhibits a single crystal basin at large $Q_6$ and a single liquid basin at small $Q_6$.   The only significant differences between the two surfaces in Panels (a) and (b) are in the location of the liquid basin and the relative stability of the crystal.  The density for the liquid basin in Panel (b) is higher than that in Panel (a), and the crystal stability in Panel (b) is reduced from that in Panel (a).  Reasons for these differences will be discussed shortly. 

\textbf{The ST2c model.}  Finally, Panel (c) shows our results for the ST2 model using the variant described in Refs. \onlinecite{sciortino2011study, poole2013free}, which we identify as the ST2c model.  It uses a reaction-field treatment of electrostatic interactions and a Lennard-Jones tail-correction.   This Panel (c), plus Panels (a) and (b) together with the results of Paper I show that small changes in temperature and variation in boundary condition have only marginal effects on the phase behavior of ST2 water.

When effects of changing potential energy or temperature are marginal, the nature of those effects are easily interpreted and computed with the identity\cite{chandler1987introduction} 
\begin{equation}
\label{eq:perturbation}
\Delta F(\rho, Q_6) =\,-\, \kB \,\ln  \langle \exp[-\Delta \mathcal{H}/\kB ]\rangle_{\rho, Q_6}\,.
\end{equation}
Here, $\Delta F(\rho,Q_6)$ is the change in free energy surface due to changing the temperature or Hamiltonian by the amount $\Delta \mathcal{H}/ \kB $.  The angle brackets refer to the ensemble average over configurations with the original temperature and Hamiltonian, and with the order parameters fixed at the values indicated by the subscripts.  When $\Delta \mathcal{H}/\kB$ is large, or when its effects are large, calculations with this formula will yield poor estimates of $\Delta  F(\rho,Q_6)$ because exponential averages are slowly converging.~\cite{pohorille2010good} On the other hand, if $\Delta \mathcal{H}$ is small, or if its effects are small, averages converge reasonably quickly.  In that case, Eq.~\ref{eq:perturbation} becomes a computationally convenient estimator.  We have checked explicitly that this measure of marginal behavior is satisfied with respect to the above cited variations for the ST2 model at $T\approx 230$~K and $N \approx 200 $.  For example, the typical size of $\Delta \mathcal{H}/\kB$ for the comparison between Figs.~\ref{Fig2} (a) and \ref{Fig5} (b) is 30\% of the root-mean-square fluctuations in the net energy per $\kB$. 

The conducting boundary condition for Ewald sums, used for Fig.~\ref{Fig5}(a), strictly cancels electrostatic surface potentials in the energy function.  These surface potentials arise from instantaneous polarization fluctuations and transient dipole moments of the total system. The conducting boundary condition is generally chosen for use with molecular dynamics simulations because the alternative, insulating boundary condition, typically introduces discontinuities in the potential energy whenever a molecule crosses the the periodic boundary.~\cite{herce2007electrostatic} For a dipole disordered system, like liquid water and ice Ih, this boundary condition should be irrelevant as its energy will average to zero.

Assuming the Ewald parameters are chosen such that energy is well converged in both cases, the pressure of the system with insulating boundary conditions is larger than that with conducting boundary conditions by an amount $\Delta p_\mathrm{surf}$.  This pressure difference is given by\cite{smith1994calculating}
\begin{equation}\label{eq:p}
\Delta p_\mathrm{surf} = \frac{2 \pi \left (2\epsilon -1 \right)}{3 \left (2 \epsilon +1 \right )} \frac{M^2}{V^2}\,.
\end{equation}
Here, $\epsilon $ is the dielectric constant of the surrounding medium, $M$ is the total system dipole, and $V$ is the volume. For a disordered system, the average $M$ is zero in the thermodynamic limit, and $M^2$ fluctuates with values that are a extensive in system size.  Accordingly, Eq. 5 shows that this strictly positive contribution to the pressure vanishes as $\sim 1/V$.  
More discussion on subtleties involved in the implementation of the Ewald summation is given in Appendix A. 

The non-electrostatic part of the ST2 potential is a 6-12 Lennard-Jones interaction.  The simulations yielding Panel (a) truncate and shift this potential to zero beyond the oxygen-oxygen distance $r_\mathrm{c}=7.5$~\AA.  This truncation produces a net potential energy that is slightly higher than that with no truncation.  An accurate correction to the potential energy that accounts for the neglected tail is the mean-field estimate $- 16\pi \epsilon_\mathrm{LJ}\sigma^6 \rho N / 3 r_\mathrm{c}^3$, where the Lennard-Jones energy and length parameters are $\epsilon_\mathrm{LJ}$ and $\sigma$, respectively.  Differentiation with respect to volume thus gives a tail correction for the pressure,\cite{Stillinger:1974p1545}
\begin{equation}
\Delta p_\mathrm{tail} = - \frac{16\pi \epsilon_\mathrm{LJ} \sigma^6 \rho^2}{3r_\mathrm{c}^3} \, .
\end{equation}
Thus, while the two simulations yielding Panels (a) and (b) in Fig.~\ref{Fig2} are both carried out at $p=2.2$ kbar, the effective pressure of the former differs from the latter by the amount $\Delta p = \Delta p_\mathrm{surf}  +  \Delta p_\mathrm{tail}$, so that the mean density of the latter will be higher than that of the former by the amount 
\begin{equation}
\Delta \rho \approx \Delta p \left[ \frac{N}{\kB}\,\frac{\langle (\delta \rho)^2 \rangle}{ \langle \rho \rangle^2} \right] \,,
\end{equation}  
where the term in square brackets is the compressibility, $(\partial \langle \rho \rangle / \partial p)_T$.  Evaluating $\Delta p$ from the formulas above for the surface and tail corrections, and estimating the mean-square density fluctuations from the widths of the liquid basins in the free energy surfaces, we find $\Delta \rho \approx 0.1$ g/cc, in harmony with the differences seen in Panels (a) and (b) of Fig.~\ref{Fig5}.  The system is much less compressible in the crystal basin than in the liquid basin (i.e., the density fluctuations are smaller in the crystal than in the liquid), and as a result, the shift in position of that basin between Panels (a) and (b) is much less than that found for the liquid. 

The relative stability of the crystal basin in Panels (b) is notably less than that in Panel (a).  This juxtaposition is another manifestation of the fact that with a given external pressure $p$, the effective pressure of the ST2b model is higher than that of the ST2a model. 

In the reaction-field treatment used to compute $F(\rho,Q_6)$ of Fig.~\ref{Fig5}(c), Coulomb interactions are summed directly up to a cutoff distance, $R_\mathrm{c}$, and contributions from larger separations are approximated as those from an ideal polarizable continuum. With this approximation, and assuming the medium has a large dielectric constant, the term 
\begin{equation}
\label{eq:DeltaU}
\Delta U_\mathrm{E} = - \frac{1}{2} \sum_{i=1}^N \sum_{j=1}^N \frac{\mathbf{\mu}_i \cdot \mathbf{\mu}_j}{R_\mathrm{c}^3}\left [1-\Theta \left(r_{ij}-R_\mathrm{c} \right) \right ] \, ,
\end{equation}
must be added to the potential energy evaluated with the truncated direct Coulomb sums.  Here, $\mathbf{\mu}_i$ is the dipole of molecule $i$, $r_{ij}$ is the distance between molecules $i$ and $j$, and the $\Theta$-function is unity for positive arguments and zero for negative arguments. This reaction-field approximation is reasonable for a homogeneous system,\cite{steinhauser1982reaction} and the asymptotic large dielectric assumption is isomorphic to the conducting boundary condition used with Ewald sums to construct Panel (a).  While, the potential energy computed with a reaction field method is not guaranteed to be the same as that computed by an Ewald summation, judicious choice of cutoff in this instance results in reasonable agreement. Indeed, there is closer correspondence between Figs.~\ref{Fig5} (a) and (c) than between Figs.~\ref{Fig5} (a) and (b).

The free energy surfaces at pressures other than those considered in Fig.~\ref{Fig5} are easily produced by re-weighting, Eq.~\ref{eq:reweight}.  Such re-weighted surfaces are shown in Appendix D.

\begin{figure}
\begin{center}
\includegraphics[width= 8cm]{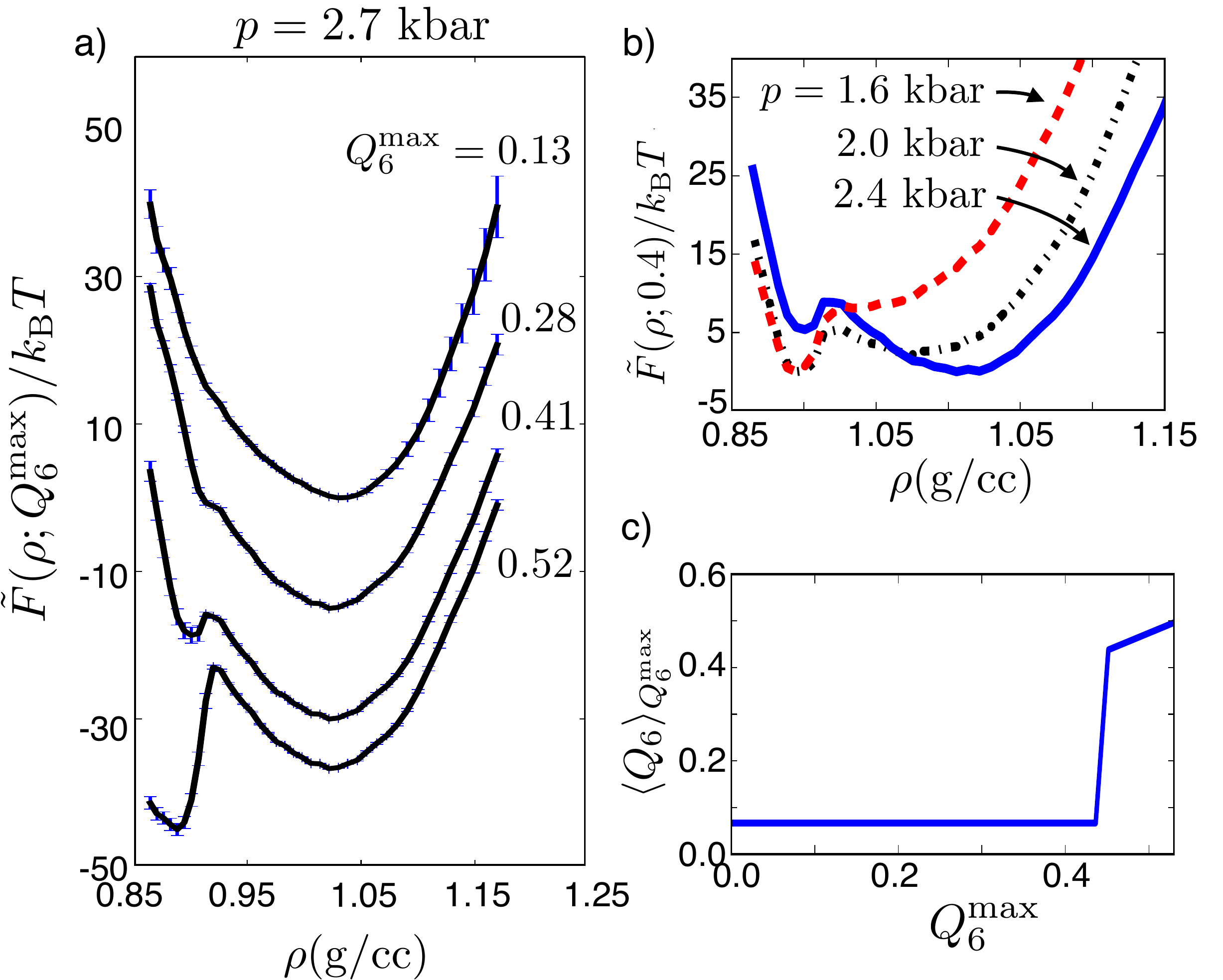}.pdf
\caption{Free energy functions and mean order parameters computed for the ST2a model at $T =235$ K, illustrating artificial polyamorphism arising from incomplete knowledge of $Q_6$-dependence in $F(\rho, Q_6)$.  (a) The restricted contracted free energy function, $\tilde{F}(\rho;Q_6^\mathrm{max})$, for $N=216$ at several indicated choices of $Q_6^\mathrm{max}$.  The functions are computed from integrating the reversible free energy function in Fig.~\ref{Fig2} (a) re-weighted to pressure $p=2.7$~kbar.  Error bars indicate one standard deviation.  Higher or lower pressures shift the free energy to favor the liquid or crystal, respectively.  See Eq.~\ref{eq:reweight}. (b) Re-weighted contracted free energy function, with $Q_6^{\mathrm{max}} = 0.4$, as if the re-weighting coincided with a Maxwell construction for two coexisting liquid phases. (c) The mean value of $Q_6$ as a function of the maximum order-parameter value for $T=235$~K and $p=2.7$~kbar. }
\label{Fig6}
\end{center}
\end{figure}
\subsection{Global order contraction~\cite{Limmer:2011arxiv} }
Here, we explicitly demonstrate the importance of $Q_6$ for analyzing phase behavior of supercooled water by showing the effects of controlling the range of accessible $Q_6$ fluctuations.  This discussion follows closely that provided in Ref.~\onlinecite{Limmer:2011arxiv}, and it augments the results of Sec. II.

In particular, we define a contraction of a constrained free energy,
\begin{equation}
\label{eq:Ftilde}
\beta \tilde{F}(\rho;Q_6^\mathrm{max}) = - \ln \left (\int_0^{Q_6^\mathrm{max}} \mathrm{d}Q_6 \, e^{-\beta  F(\rho,Q_6) } \right )\, .
\end{equation} 
We compute these functions from our estimates of the unconstrained reversible free energy surface.  Functions so obtained are shown in Fig. \ref{Fig6}(a). The unconstrained free energy from which they are derived is $F(\rho,Q_6)$ graphed in Fig.~\ref{Fig2}(a) and re-weighted to the pressure 2.7 kbar.  It is the reversible free energy surface for the ST2a model at a pressure that puts the system close to coexistence between the liquid and the crystal at the temperature $T=235\,$K.  At lower pressures, the liquid will be supercooled; at higher pressures, the liquid will be stable with respect to the crystal.  The surfaces for those different pressures are obtained from the pictured surface by applying Eq.~\ref{eq:reweight}.

An upper limit of $Q_6^\mathrm{max}=0.13$ encompasses the liquid basin.  The contracted free energy in that case is unimodal, and it does not exhibit statistically meaningful changes in convexity.  That is to say, no re-weighting with Eq.~\ref{eq:reweight} in that case will produce bi-modality.  Thus, there is not a second liquid for all densities (and corresponding pressures) in the range considered. 

\begin{figure}
\begin{center}
\includegraphics[width= 8cm]{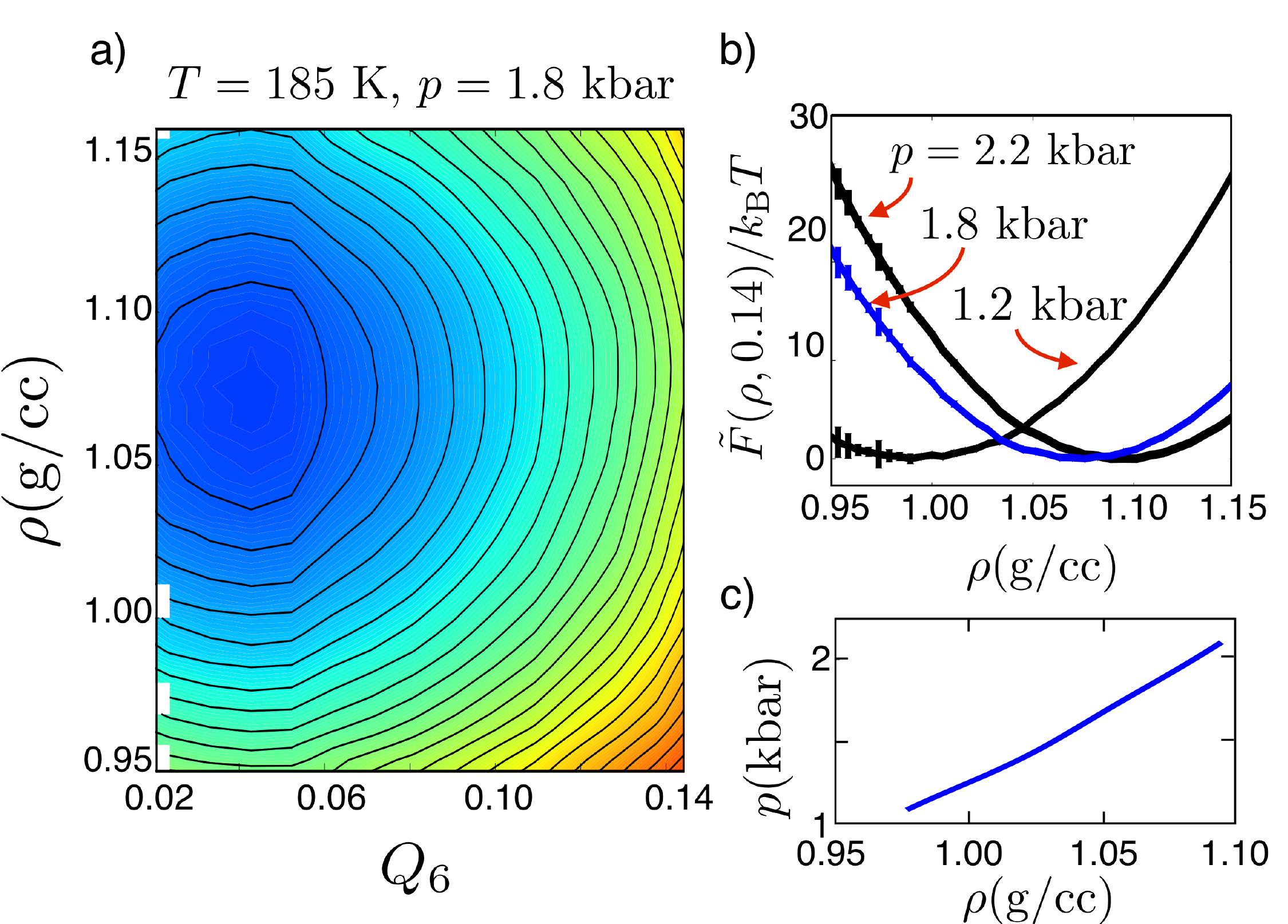}
\caption{Reversible free energies and equation of state for liquid TIP4P/2005 at $T=185$K and $N=216$ for pressures between 1 and 3 kbar.  a) Free energy, $F(\rho,Q_6)$. Contour lines are separated by 1~$\kB$ and error estimates are less than 1~$\kB$.  b) Contracted free energy as a function of density when restricting $Q_6$ to the liquid basin, i.e., $Q_6 < 0.14$.  c) Liquid phase equation of state accessible from the free energy surfaces shown in Panels (a) and (b)  }
\label{Fig7}
\end{center}
\end{figure}
Notice, however, that this contracted free energy function is skewed in a fashion where fluctuations towards low density are more probable than fluctuations towards higher density.  In the limit of large system size, fluctuations within stable or metastable basins are Gaussian.  The skewed behavior is therefore a finite system-size effect. Its physical origin can be resolved by increasing $Q_6^\mathrm{max}$. For $Q_6^\mathrm{max}=0.25$, a shoulder and change in convexity appears.  For larger values of $Q_6^\mathrm{max}$, there is systematic growth of the shoulder into a basin. This low density basin is the crystal.  The mean value of $Q_6$ as a function of  $Q_6^\mathrm{max}$, $\langle Q_6 \rangle_{Q_6^\mathrm{max}}$, is shown in Fig. \ref{Fig6}(c).  This mean value remains at its liquid state value for  $Q_6^\mathrm{max}<0.45$. Thus, by sampling the full surface pictured in Fig.~\ref{Fig2}(a), it is possible to identify the low-density basin as the crystal phase.  

\begin{figure*}[t]
\begin{center}
\includegraphics[width= 15cm]{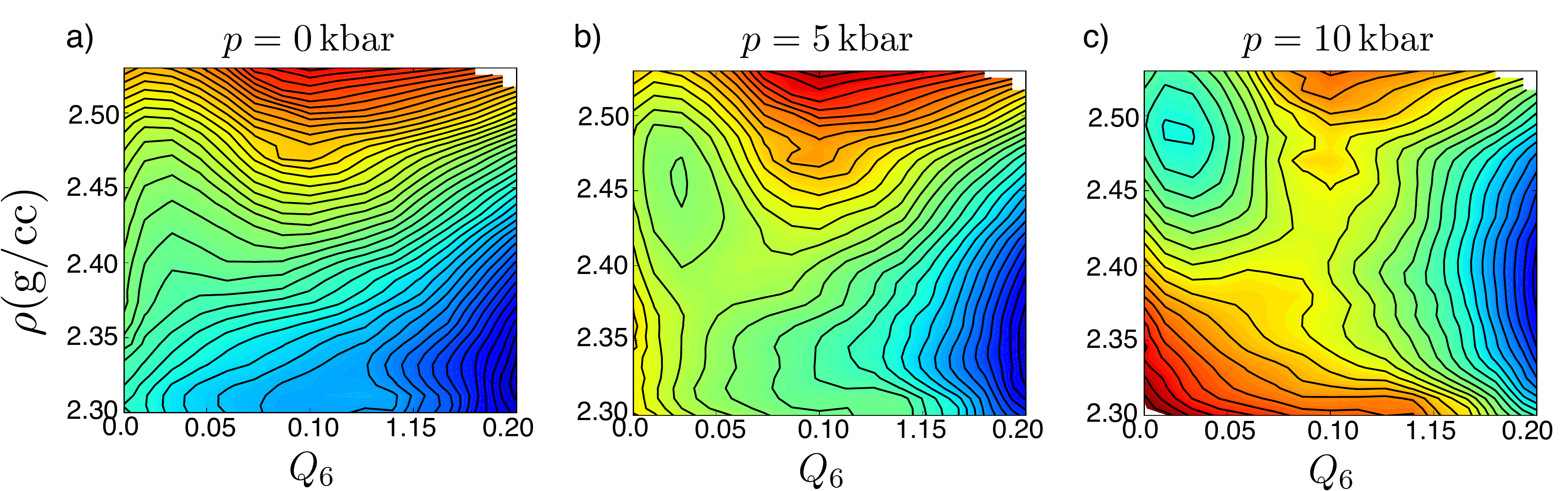}
\caption{Reversible free energy, $F(\rho,Q_6)$, computed for the SW model for $N=512$, $T=1050$~K and three pressures.  Contour lines are separated by 2 $\kB$ and error estimates are less than 1 $\kB$.}
\label{Fig8}
\end{center}
\end{figure*}
When limiting the range of $Q_6$ to  $Q_6^\mathrm{max}<0.45$, the features shown in Fig. \ref{Fig6} could be easily misinterpreted as indicative of liquid-liquid coexistence.  Indeed, by applying an external pressure to re-weight the curves in Fig.~\ref{Fig6} (a), with Eq.~\ref{eq:reweight} as if $\tilde{F}(\rho, Q_6^\mathrm{max})$ was an equilibrium free energy for $0.13<Q_6^\mathrm{max}<0.43$, a bistable density distribution can be obtained with a low mean value of $Q_6$.  This type of construction is illustrated in Panel (b) of Fig. \ref{Fig6}.  The bi-stability of precisely the sort reported in Refs.~\onlinecite{liu2009low,liu2012liquid,sciortino2011study, poole2013free} is thereby found.  But contrary to the interpretation expressed in Ref.~\onlinecite{poole2013free}, the behavior is not reflective of liquid-liquid transition.  Rather, the unconstrained free energy shows that the appearance of convexity loss is associated with moving towards and then over the barrier separating liquid and crystal basins.  

As the geometry of the total $F(\rho, Q_6)$ changes with system size, $N$, the re-weighting (i.e., pseudo Maxwell construction) illustrated in Panel (b) and alluded to in Ref.~\onlinecite{poole2013free} will depend upon system size in ways that are inconsistent with two-phase coexistence.  The free energy barrier separating basins of truly coexisting phases scales as $N^{2/3}$.  But the low density phase with $Q_6$ confined to low values cannot equilibrate and thus cannot coexist. 

\section{Studies of additional computer simulation models}
The potential existence of a second liquid phase has been conjectured for a number of different models of water and other tetrahedral liquids. In this section we summarize our work on applying robust free energy calculations to a selection of these models.

\subsection*{Larger phase space for the mW model}
In Paper I,\cite{Limmer:2011p134503} we reported results for the mW model over a range of temperatures spanning 160 K to 320 K and pressures spanning 1 bar to 3 kbar.  A corresponding states argument\cite{limmer2012phase} indicates that this pressure range for mW water is a factor of 3 smaller than that accessed in experiments.  As a consequence, the range of pressures relevant to the existence of a liquid-liquid transition in the mW model extends up to 10 kbar. We have thus expanded the domain of our free energy calculations and ruled out a liquid-liquid transition up to 10 kbar and $160$ K$<T<320$ K.  The free energy surfaces throughout are consistent with those shown in Paper I.

\subsection*{Quantitative water model (TIP4P/2005)}
The recently parameterized TIP4P/2005 water model\cite{abascal2005general} has had success in quantitatively reproducing many essential properties of water and ice, including the density versus temperature line at $p=1$ bar. Previous studies have extrapolated equation of state data for this model to estimate the location of a putative liquid-liquid critical point and first order transition line.\cite{abascal2010widom} Using free energy calculations we can test the validity of this extrapolation. 

Figure \ref{Fig7} shows $F(\rho,Q_6)$ for TIP4P/2005 computed at $T=185$K and $p=1.8$ kbar for $N=216$ molecules. Reference \onlinecite{abascal2010widom} estimates the location of the critical point to be $T= 193$ K and $p=1.35$ kbar, as shown in Panels (b) and (c) of Fig.~\ref{Fig7}. The free energy in Fig.~\ref{Fig7} is computed below this purported critical point temperature, yet the distribution is clearly monostable at low values of $Q_6$.  High values of $Q_6$ were sampled, but not shown for clarity. On re-weighting this free energy, Eq.~\ref{eq:reweight}, bi-stability does not appear throughout the pressure range, 1 kbar $<p<$ 3 kbar.  We can conclude, therefore, that for this model of water, a second liquid does not exist at conditions studied.  Here too, the putative liquid-liquid transition seems to be an artifact of finite-time sampling, reflecting a liquid-to-ice transition where coarsening is incomplete.

\subsection*{General tetrahedral model (SW)}
We have also studied the behavior of the Stillinger-Weber (SW) model of silicon. This model has been the subject of numerous studies\cite{sastry2003liquid,beaucage2005liquid,vasisht2011liquid} that have used equation of state data to propose the existence of a second liquid phase. Using free energy methods, we can test this proposal by examining conditions below the putative liquid-liquid critical temperature. 

Figures \ref{Fig8} (a)-(c) shows $F(\rho,Q_6)$ calculation for a system of 512 particles at $T=1050$ K and a range of pressures, 0 $<p<$ 10 kbar. Based on the phase diagram proposed in Ref.~\onlinecite{vasisht2011liquid}, also calculated with 512 particles and citing agreement with Ref.~\onlinecite{sastry2003liquid}, this temperature and pressure range should traverse the first-order liquid-liquid phase boundary. Rather than finding two liquid basins upon decreasing pressure, we find the system crosses the line of liquid stability, $T_s(p)$, for pressures lower than $p=1$ kbar.

Put into context with our discussion in Sec. II, the geometry of the free energy in Fig.~\ref{Fig8} (a) illustrates how finite-time sampling can produce the illusion of liquid-liquid bistability.  Specifically, the facile equilibration in density would result in an abrupt change between the liquid with density at $\rho=2.45$ g/cc and an amorphous material with $\rho=2.35$ g/cc, while over short timescales the slow diffusion in $Q_6$ would keep the value small. In fact this point was noted by the authors of Ref.~\onlinecite{vasisht2011liquid} when they state that out of 10 to 50 trajectories, ``Non-crystallizing samples (an average of 5) were run for up to 10 relaxation times when possible." 

\section{Summary of calculations on putative liquid-liquid transitions in supercooled tetrahedral fluids}

Table~\ref{Summary} summarizes all that is now known about the putative liquid-liquid transition in supercooled water and related systems.  We see that the low-temperature behaviors of several different models of water-like liquids are similar.  The reversible phase behavior in all cases is that of one liquid, a liquid that can coexist with and transform to a lower density ice-like phase.  In all cases, the time scales for density fluctuations in the supercooled liquid are several orders of magnitude shorter than those for long-ranged order fluctuations, and both time scales are strongly temperature dependent.  

These features lead to a rich non-equilibrium behavior, some of which we have illustrated here in our treatment of the early stages of coarsening, and previously in our treatment of confined supercooled water.\cite{limmer2012phase}  This non-equilibrium behavior is clearly responsible for the numerous reports of polyamorphism in computer simulations of water, and it is likely important in processes that lead to the formation of glassy phases of water.  While the former represent artifacts of finite-time sampling, the behaviors of glasses represent an important class of phenomena worthy of future study.  Such far-from equilibrium behavior, however, is beyond the  scope of what we have considered here.

The generic nature of what is found in so many models makes it seem unlikely that plausible models of water will exhibit liquid-liquid coexistence and a second low-temperature critical point.  It also suggests that any model exhibiting a crystal phase with lower density than the liquid phase will also exhibit transient behavior that looks like polyamorphism when viewed on the time scales no longer than those of early-stage coarsening.  The reason why this non-equilibrium behavior is not observed in a recent study of the SPC/E model\cite{Berendsen1987SPCE} is because that study\cite{giovambattista2012interplay} quenches from temperatures significantly higher than $T_\mathrm{s}$ for that model. 

The generic nature also implies that simplified models, like mW water, can reliably serve as useful descriptors of water. Hypotheses on the nature of low temperature water could often be studied in that way.  Further, computational efforts that struggle with overcoming the separation of time scales inherent in low-temperature water could often be tested with such models.  In our own work (see Appendix C below), it seems clear that errors in prior announcements of liquid-liquid transitions in supercooled water could have been detected by first examining the behavior of mW and SW models.

In documenting the necessity of attending to time scales and relaxation, we have outlined robust methods by which equilibration and reversibility can be achieved.  There is clearly need for independent assessment of this growing body of work, which we look forward to seeing in the future.

\begin{table}
  \caption{Summary of models and conditions considered in this and in previous studies}
  \label{Summary}
   \begin{tabular}{l >{\centering}m{4cm} c }
    \hline
    Model 	& 	AP\footnotemark[1]$(T/\mathrm{K},p/\mathrm{kbar})$ 	& 	FE\footnotemark[2] $(T/\mathrm{K},p/\mathrm{kbar})$ \\
    \hline\hline
    mW 				& - 								 &(160-300, 0-10.0)\footnotemark[8]\\ \\
    ST2b 	& (242, 1.8)\footnotemark[3] (240, 1.8)\footnotemark[3] \,\,\,\,\,\,(238, 1.9)\footnotemark[3] (235, 2.0)\footnotemark[3] \footnotemark[4]  (228, 2.2)\footnotemark[4] (224, 2.3)\footnotemark[4] & (230-240, 1.0-3.0)\footnotemark[9] \\ \\
    ST2c 	& (245, 1.8)\footnotemark[4] (240, 2.0)\footnotemark[4]\,\,\,\,\, (235, 2.2)\footnotemark[4] (230, 2.4)\footnotemark[4] & (230-240, 1.0-3.0)\footnotemark[9] \\ \\
    ST2a 	& -  & (230-240, 1.0-3.0)\footnotemark[9] \\ \\
    SW 				& (1070, 0.)\footnotemark[6]  (950, 7.5)\footnotemark[6] \,\,\,\,\,(920, 11.3)\footnotemark[6] & (1050, 0-10.0)\footnotemark[10] \\ \\
    TIP4P/2005 			& (195, 1.45)\footnotemark[7] & (180-190, 1.0-2.6)\footnotemark[9] \\ \\
    \hline \hline 	       
  \end{tabular}
\footnotetext[1]{Conditions where artificial polyamorphism has been reported.}
\footnotetext[2]{Conditions where $Q_6$-equilibrated free energy calculations rule out a liquid-liquid transition.}
\footnotetext[3]{Short Grand Canonical simulations from Ref.~ \onlinecite{liu2009low}, $N\approx 200$.}
\footnotetext[4]{$Q_6$-unequilibrated free energy calculations from Ref.~\onlinecite{liu2012liquid}, $N\approx 200$.}
\footnotetext[5]{$Q_6$-unequilibrated free energy calculations from Ref.~ \onlinecite{poole2013free}, $N\approx 200$}
\footnotetext[6]{Short molecular dynamics trajectories from Ref.~\onlinecite{vasisht2011liquid}, $N= 512$.} 
\footnotetext[7]{Short molecular dynamics trajectories from Ref.~\onlinecite{abascal2010widom}, $N= 500$. }
\footnotetext[8]{This paper and Paper I, $N=216$ to $N=1000$. }
\footnotetext[9]{This paper, $N=216$. }
\footnotetext[10]{This paper and Paper I, $N=512$. }
\end{table}

\begin{figure}[t]
\begin{center}
\includegraphics[width= 8cm]{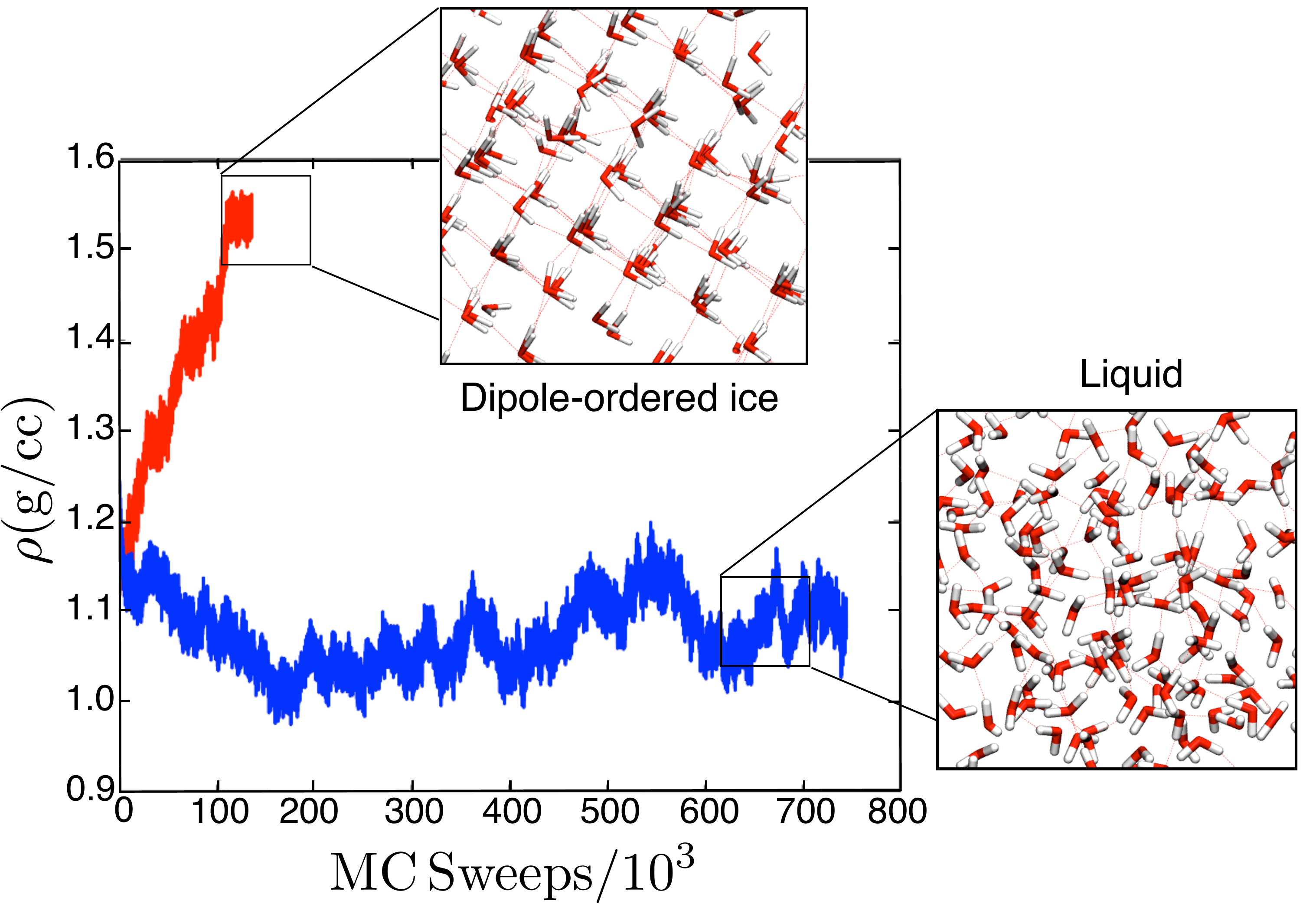}
\caption{Time series illustrating how some choices of parameters for the Ewald sum in the ST2a model, which uses conducting boundary conditions, can lead to the formation of dipole ordered ice, while for other choices the liquid remains stable. See text. }
\label{Fig9}
\end{center}
\end{figure}

\begin{acknowledgments}
We thank Phillip Geissler, Alex Hudson, Aaron Keys, Milo Lin and Dayton Thorpe for comments on an earlier version of the manuscript.  We have also benefitted from discussion with the authors of Refs.~\onlinecite{liu2012liquid} and \onlinecite{poole2013free}.  Early work on this project was supported by the Director, Office of Science, Office of Basic Energy Sciences, Materials Sciences and Engineering Division and Chemical Sciences, Geosciences, and Biosciences Division of the U.S. Department of Energy under Contract No. DE-AC02-05CH11231. While for the later stages the authors gratefully acknowledge the Helios Solar Energy Research Center, which is supported by the Director, Office of Science, Office of Basic Energy Sciences of the U.S. Department of Energy under Contract No. DE-AC02-05CH11231.\end{acknowledgments}
\begin{figure*}
\begin{center}
\includegraphics[width= 15cm]{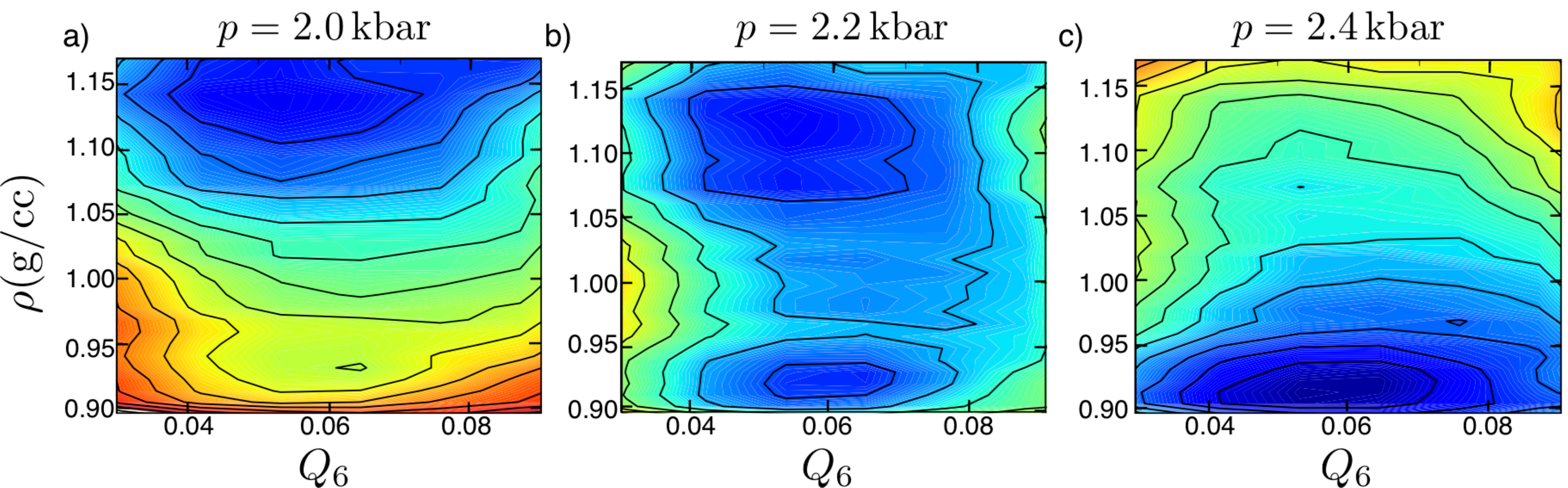}
\caption{ Nonequilibrium pseudo free energy surfaces, $F_\mathrm{ne}(\rho, Q_6, t)$, at three different pressures, illustrating how artificial polyamorphism arises as a finite-time effect.  All three surfaces are evaluated by propagating from an initial liquid distribution for a time $t = 10 \tau_{Q_6}$.  Computed for the ST2b model from the Fokker-Planck analysis with the underlying reversible free energy surface shown in Fig.~\ref{Fig5}{b}.  The appearance of artificial polyamorphism is like that found in Ref.~\onlinecite{liu2012liquid} for similar conditions.  The time $t = 10\,\tau_{Q_6}$ corresponds to the time used for averaging in Ref.~\onlinecite{liu2012liquid}. Contour lines are separated by 1~$\kB$ and statistical uncertainties are less than 1~$\kB$. The correct, reversible free energy surface for $p=2.2$~kbar is shown in Fig.~\ref{Fig5}(b).}
\label{Fig10}
\end{center}
\end{figure*}

\appendix

\section{Checks on coding and long-ranged force evaluations}
While the main text has postulated reasoned explanations for a discrepancy between the calculations of our pervious work and that of Ref.~\onlinecite{liu2009low,sciortino2011study,liu2012liquid,poole2013free}, it does not address the possibility of an undetermined error in implementation. To address this we have worked with the authors of Ref.~\onlinecite{liu2012liquid} in exchanging data and molecular configurations.\footnote{We are grateful to Pablo Debenedetti,  Yang Liu, Jeremey Palmer, Athanassios Panagiotopoulos for sharing data and results with us.} The results of these studies were unambiguous: all parties were able to reproduce energy evaluations and standard mean and fluctuation quantities for the ST2b model. Specific statistical properties computed were the average density and compressibility for a system of 200 molecules at two state points, $T=300$~K, $p=1$ bar and $T=235$~K, $p=2.2$~kbar.  Agreement between calculations was found to be within statistical error.\footnote{Y. Liu, Private communication 2012} Moreover, tests were done swapping configurations of $N=200$ molecules and obtaining independent evaluation of the energy and $Q_6$ values.  These tests demonstrated that both groups were evaluating the same energy function and the same order parameter. Thus, an implementation error does not seem to be the root cause of this discrepancy.

\begin{figure}[b]
\begin{center}
\includegraphics[width= 8cm]{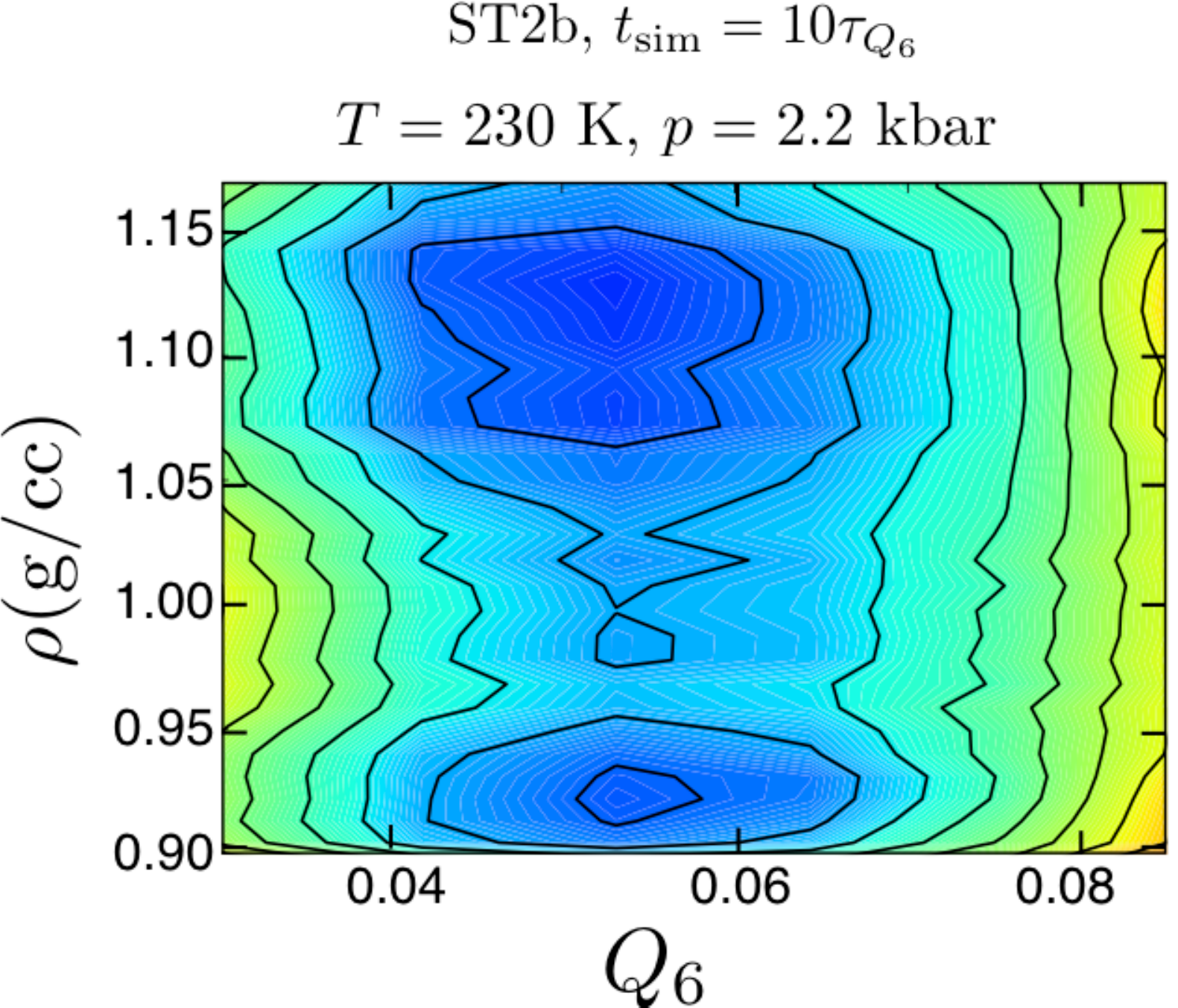}
\caption{Nonequilibrium pseudo free energy surfaces, $F_\mathrm{ne}(\rho, Q_6, t)$ illustrating how artificial polyamorphism arises as a finite-time effect.  The correct reversible surface is shown in Fig.~\ref{Fig5}(b). The incorrect surface shown here is computed for the ST2b model from a non equilibrium distribution for $Q_6$, where the distribution is obtained from simulations initiated in the liquid basin and propagated for a time $10\, \tau_{Q_6}$. The surface shows the appearance of artificial polyamorphism like that found in Ref.~\onlinecite{liu2012liquid}  for similar conditions.  The simulation time $t_\mathrm{sim} = 10\,\tau_{Q_6}$ corresponds to the time used for averaging in Ref.~\onlinecite{liu2012liquid}. Contour lines are separated by 1 $\kB$ and statistical uncertainties are less than 1 $\kB$.}
\label{Fig11}
\end{center}
\end{figure}
As reported in Ref.~\onlinecite{liu2012liquid}, at specific regions of state space, simulations of a stable low temperature liquid are sensitive to details of the Ewald summation. Specifically, the authors of Ref.~\onlinecite{liu2012liquid} found that with conducting boundary conditions and $N\approx 200$, liquid configurations spontaneously evolve into a dipole ordered form of ice VII when pressure is elevated and $T<235$ K.  While the authors of that study report they were unable to find Ewald parameters under conducting boundary conditions that did not display this pathological behavior, other studies applying this boundary condition with other electrostatic estimators have not found this same difficulty.  For example, Ref.~\onlinecite{poole2013free} employs a reaction-field treatment of electrostatic interactions with conducting boundary conditions.

By juggling the method of evaluating Ewald sums, we too have been able to reproduce the pathological behavior reported in Ref.~\onlinecite{liu2012liquid}.  Two of our time series are shown in Fig.~\ref{Fig9}.
Using hybrid MC dynamics for 216 molecules and conducting boundary conditions at $T=228$ K and $p=2.2$ kbar, we find that changes to the details of the Ewald parameters can result in either stable liquid behavior (blue curve), or spontaneous dipole ordering (red curve). In the stable liquid case, the Ewald parameters are chosen to accurately estimate the energy and forces of the long range part of the potential to 1 part in $10^4$ using a standard error estimator and a spherical wave-vector cutoff.~\cite{kolafa1992cutoff} In order to maintain this level of accuracy with a changing box size the parameters are updated over the course of the trajectory. In the unstable trajectory a cubic wave-vector cutoff is used, and the parameters are held fixed which reduces the accuracy of the potential estimate. There may be other ways to induce this pathological behavior, and these ways will depend upon the type of dynamics used.  The principal point is that it is a finite-size effect that is sensitive to technical but unphysical details in implementing Ewald sums, and the effect can be avoided when sufficient care is applied to the algorithm. 

\section{Early stage coarsening and artificial polyamorphism in the ST2b model}
The calculations carried out in Section II for the ST2a model have been also done for the ST2b model at $T=230$ K with $N=216$ and pressures ranging from 2 kbar to 2.4 kbar.  These are the conditions and model considered in Ref.~\onlinecite{liu2012liquid}.  The results are shown in Fig.~\ref{Fig10}.  The theoretical results agree with those found in Ref.~\onlinecite{liu2012liquid}, thus indicating that the free energies reported in that work suffer from finite-time effects.

In addition to the Fokker-Plank analysis, we have computed $P_\mathrm{ne}(Q_6,t)$ directly from a molecular simulation of the ST2b model, starting from an ensemble of liquid configurations and running for $t_\mathrm{sim}=10 \tau_{Q_6}$.  We have then convoluted that non-equilibrium distribution with the equilibrium $P(\rho | Q_6)$ obtained from the reversible free energy in Fig.~\ref{Fig5}(b) according to Eq.~\ref{eq:factorization1}.  The results are shown in Fig.~\ref{Fig11}.  Again, behavior like that reported in Ref.~\onlinecite{liu2012liquid} is obtained, thus indicating that the results of that paper suffer from finite-time effects.

\section{Artifacts of un-equilibrated initial conditions}
\begin{figure}
\begin{center}
\includegraphics[width= 8cm]{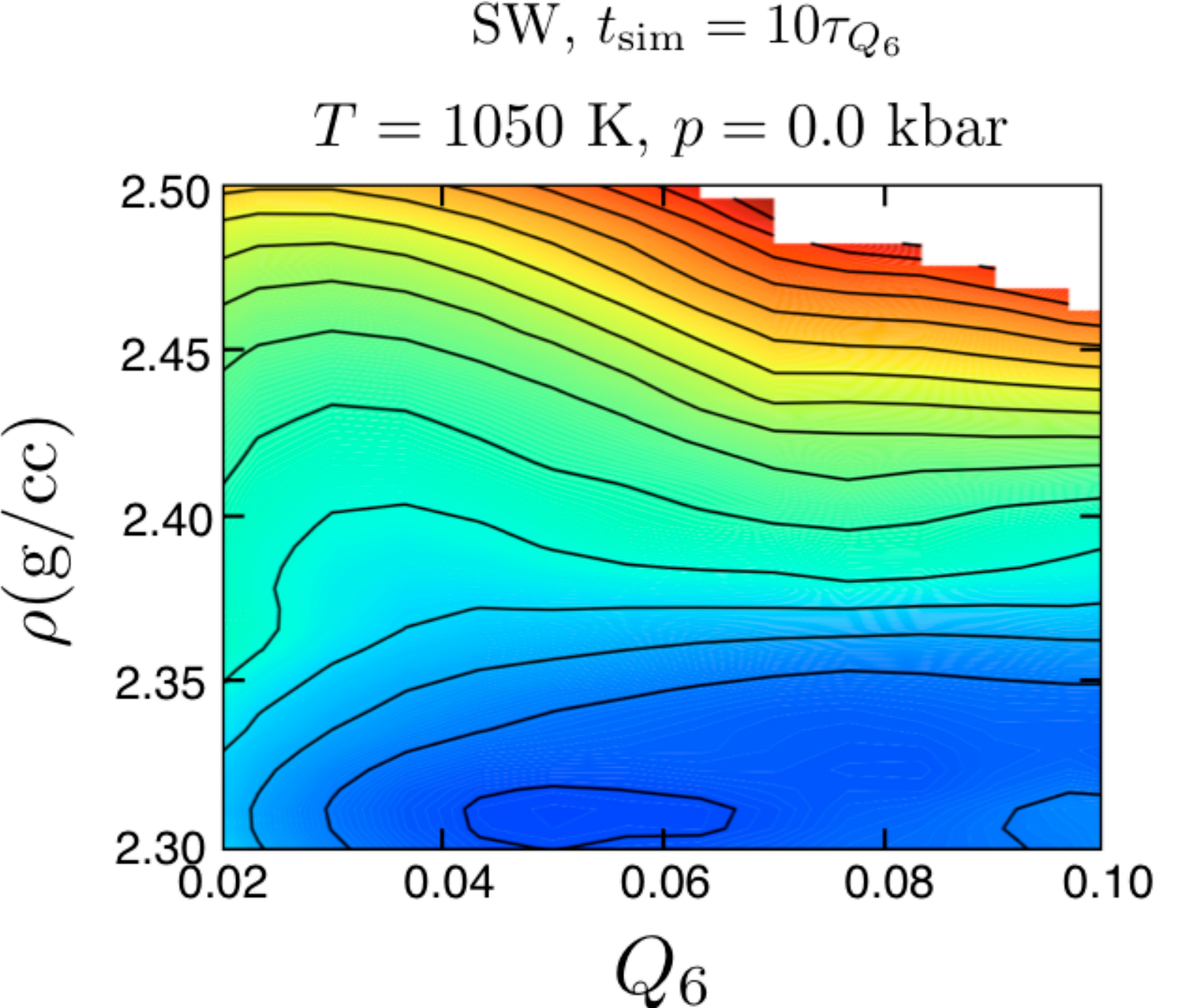}
\caption{Nonequilibrium pseudo free energy surface for the SW model obtained with initial conditions from $T=1100$~K and short equilibration times. The correct reversible surface is shown in Fig.~\ref{Fig8}(a). Contour lines are separated by $3~\kB$ and statistical uncertainties are about  1 $\kB$.}
\label{Fig12}
\end{center}
\end{figure}

A somewhat different but related source of systematic error would be found in simulations that do not completely equilibrate from initial configurations taken from a higher temperature.  For example, suppose one has established good equilibrium statistics for a system at temperature $T+\Delta T$, and then wishes to use configuration from that data to seed a free energy calculation at the temperature $T$.  (One can consider parallel tempering\cite{frenkel2001understanding} as one realization of this idea.)  If at $T+\Delta T$ the system exists in a metastable liquid that becomes unstable at the lower temperature $T$, then configurations from the higher temperature will bias the statistics of the lower-temperature system for time scales short compared to average time for $Q_6$ to equilibrate (i.e., to leave the liquid region).  Therefore, if the runs performed at temperature $T$ are too short, a remnant of the meta-stable liquid basin at temperature $T+\Delta T$ will remain in the estimate of the free energy at temperature $T$.  Figure~\ref{Fig12} illustrates this problem for the case of the SW model.

The correct reversible free energy surface for the SW model at this condition is shown in Fig.~\ref{Fig8}(c).  In Fig.~\ref{Fig12}, we show an incorrect surface, which is obtained by using equilibrated data at temperature $T+\Delta T = 1100$ K as initial conditions, and then carrying out trajectories at $T=1050$ K that run for only $t_\mathrm{sim} =10 \tau_{Q_6}$.  Histograms produced by this procedure yield a pseudo free energy with a low-density liquid minimum at conditions where the actual liquid is thermodynamically unstable.  Error estimates performed from the data collected in this way are very small, of the order of 1 $\kB$.  These small error estimates reflect the highly correlated nature of the data.  Changes of $Q_6$ are simply too slow to fully develop on the time scales probed.

\section{Pressure dependence of $F(\rho, Q_6)$ for variants of the ST2 model}
\begin{figure*}[t]
\begin{center}
\includegraphics[width= 15cm]{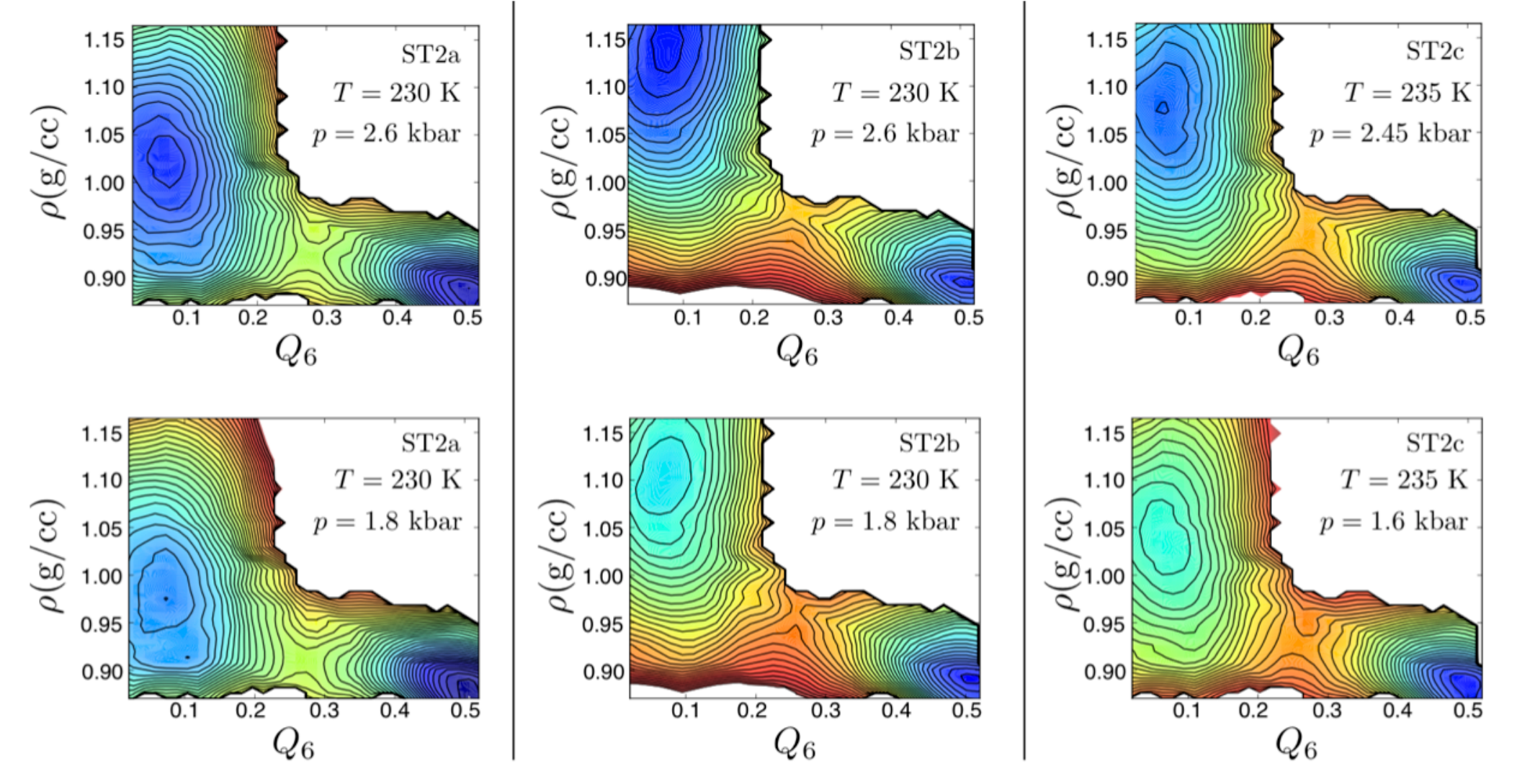}
\caption{Pressure variation of free energy surfaces, $F(\rho,Q_6; p,T )$, for three variants of the ST2 model at temperatures where others report evidence of liquid-liquid coexistence for the ST2 model.  Surfaces at intermediate pressures are shown in Fig.~\ref{Fig5}.  Contour line spacing and system size are the same as those in Fig.~\ref{Fig5}. }
\label{Fig13}
\end{center}
\end{figure*}

References~\onlinecite{liu2012liquid} and \onlinecite{poole2013free} suggest that the analysis of the ST2a model in Paper I~\cite{Limmer:2011p134503} overlooks a liquid-liquid transition because it examines a pressure that lies outside a hypothesized spinodal region.  We examine the validity of this suggestion with Fig. \ref{Fig13}.  This figure shows the free energy surfaces at different pressures for the three variants of the ST2 considered in the main text.  The pressure variations are constructed by applying Eq.~\ref{eq:reweight} to the surfaces graphed in Fig.~\ref{Fig5}.  What is found in each case is that pressures higher than those considered in Fig.~\ref{Fig5} shift stability towards the liquid and increases density of the liquid, and pressures lower than those considered in Fig.~\ref{Fig5} shift stability towards the crystal and decreases density of the liquid.  Variation of pressure over the ranges considered by Refs.~\onlinecite{liu2012liquid} and \onlinecite{poole2013free} does not lead to liquid bi-stability in the reversible behavior of the ST2 model.  There is one liquid phase, and no liquid-liquid coexistence or spinodal.

%

\end{document}